\documentclass[prd,superscriptaddress,altaffilletter,showpacs,nofootinbib]{revtex4}

\usepackage{amssymb}
\usepackage[dvips]{graphicx}
\usepackage{amsmath}

\newcommand{\be}{\begin{equation}}
\newcommand{\ee}{\end{equation}}
\newcommand{\bea}{\begin{eqnarray}}
\newcommand{\eea}{\end{eqnarray}}

\newcommand{\der}{\partial}
\newcommand{\vphi}{\varphi}

\begin{document}

\title{On the physical consequences of a Weyl invariant theory of gravity}

\author{Israel Quiros}\email{iquiros@fisica.ugto.mx}\affiliation{Departamento de Ingenier\'ia Civil, Divisi\'on de Ingenier\'ia, Universidad de Guanajuato, Gto., M\'exico.}

\date{\today}

\begin{abstract} In this paper we explore the physical consequences of assuming Weyl invariance of the laws of gravity from the classical standpoint exclusively. Actual Weyl invariance requires to replace the underlying Riemannian geometrical structure of the background spacetimes by Weyl integrable geometry (WIG). We show that gauge freedom, a distinctive feature of Weyl invariant theories of gravity, leads to very unusual consequences. For instance, within the cosmological setting in a WIG-based conformal invariant gravity theory, also known as conformal general relativity (CGR), a static universe is physically equivalent to a universe undergoing de Sitter expansion. It happens also that spherically symmetric black holes are physically equivalent to singularity-free wormholes. Another outstanding consequence of gauge freedom in the framework of CGR is that inflation is not required to explain the flatness, horizon and relict particle abundances, among other puzzles that arise in standard GR-based cosmology. Besides, the cosmological constant and the mass hierarchy problems do not arise neither in this setup.\end{abstract}


\maketitle


\section{Introduction}\label{intro} 

Local scale transformations and Weyl rescalings are of great importance in quantum field theory \cite{nakayama_review} and in the study of the fundamental interactions of fields. These are required for the renormalization procedure to work appropriately at very short distances \cite{smolin_npb_1979, zee_prd_1981, cheng_prl_1988, goldberger_prl_2008}. In Ref. \cite{percacci} it is shown how to construct renormalization group equations for Weyl invariant theories, while in Ref. \cite{odintsov, odintsov_1, odintsov_2, odintsov_3} it has been shown that scale invariance is very much related with the effect of asymptotic conformal invariance, where quantum field theory predicts that theory becomes effectively conformal invariant.  

Within the framework of gravitational theories Weyl invariance has been investigated from different perspectives \cite{smolin_npb_1979, zee_prd_1981, cheng_prl_1988, goldberger_prl_2008, percacci, odintsov, odintsov_1, odintsov_2, odintsov_3, utiyama, indios, indios_1, shapo, prester, padilla, bars, bars_1, bars_2, carrasco, quiros_1, jackiw, alpha, alpha_2, alpha_3, alpha_4, salvio-1, ghorbani, farz, vanzo, alvarez, alvarez-jcap-2015, khoze, karananas, tambalo, kannike, javier, ferreira, maeder, myung, moon}. In Ref. \cite{weyl} the first serious attempt to create a scale-invariant theory of gravity (and of electromagnetism) was undertaken. This attempt was unsuccessful due to an unobserved broadening of the atomic spectral lines \cite{scholz_1, scholz_2, perlick, novello}. A scale-invariant extension of general relativity (GR) based on Weyl's geometry was explored in Ref. \cite{smolin_npb_1979}. If the theory contains a Higgs phase then, at large distances this phase reduces to GR. In Ref. \cite{zee_prd_1981} it has been shown that gravity may arise as consequence of dynamical symmetry breaking in a scale (also gauge) invariant world. A quantum field theory of electroweak (EW) and gravitational interactions with local scale invariance and local $SU(2)\times U(1)$ gauge invariance was proposed in Ref. \cite{cheng_prl_1988}. The requirement of local scale invariance leads to the existence of Weyl's vector meson which absorbs the Higgs particle remaining in the standard model of particles (SMP). 

In Refs. \cite{bars, bars_1, bars_2} it is shown how to lift a generic non-scale invariant action in Einstein frame into a Weyl invariant theory and a new general form for Lagrangians consistent with Weyl symmetry is presented. Advantages of such a conformally invariant formulation of particle physics and gravity include the possibility of constructing geodesically complete cosmologies \cite{bars_1}. In this regard see critical comments in Refs. \cite{carrasco, quiros_1, jackiw} and the reply in Ref. \cite{bars_2}. In Refs. \cite{alpha} a new class of chaotic inflation models with spontaneously broken conformal invariance has been developed. In this vein a broad class of multi-field inflationary models with spontaneously broken conformal invariance is described in Ref. \cite{alpha_2}, while generalized versions of these models where the inflaton has a non-minimal coupling to gravity with $\xi<0$ ($\xi\neq-1/6$), are investigated in Ref. \cite{alpha_3}.

Weyl invariant theories of gravity are usually associated with Lagrangian density terms that are quadratic in the Weyl tensor, $C_{\mu\nu\sigma\lambda}$: ${\cal L}=\sqrt{|g|}\,C_{\mu\nu\sigma\lambda}C^{\mu\nu\sigma\lambda},$ or that contain other Weyl invariant terms like: $\sqrt{|g|}\,R^2$, or like $\sqrt{|g|}\,R_{\mu\nu}R^{\mu\nu}$, where $R$ and $R_{\mu\nu}$, are the Ricci curvature scalar and tensor, respectively. Among the problems of these theories, being based in higher-order Lagrangians, is the one associated with the Ostrogradsky instability and with the presence of other ghosts \cite{ostro-theor, woodard_2007, stelle, ovrut}. Another way in which any theory of gravity can be made Weyl invariant is by introducing a conformally coupled scalar field like in the following action:

\bea S=\int d^4x\sqrt{|g|}\left[\frac{\phi^2}{12}\,R+\frac{1}{2}(\der\phi)^2-\frac{\lambda}{12}\,\phi^4\right].\label{deser-action}\eea Notice that under the following redefinition of the scalar field, $\phi^2\rightarrow 6\exp\vphi$, the action \eqref{deser-action} can be written in terms of the so called ``string frame'' variables: $$S=\frac{1}{2}\int d^4x\sqrt{|g|}\,e^\vphi\left[R+\frac{3}{2}(\der\vphi)^2-3\lambda\,e^\vphi\right].$$ This is actually Brans-Dicke theory with the singular value of the coupling parameter $\omega_\text{BD}=-3/2$.

Under the Weyl rescalings:

\bea g_{\mu\nu}\rightarrow\Omega^{-2}g_{\mu\nu},\;\phi\rightarrow\Omega\,\phi,\label{scale-t}\eea where the non-vanishing positive smooth function $\Omega^2(x)$ is the conformal factor, the combination $\sqrt{|g|}[\phi^2R+6(\der\phi)^2]$ as well as the scalar density $\sqrt{|g|}\,\phi^4$, are kept unchanged. Then, if further assume that the dimensionless constant $\lambda$ is not transformed by the Weyl rescalings, the action (\ref{deser-action}) is invariant under \eqref{scale-t}. Any scalar field which appears in the gravitational action the way $\phi$ does in \eqref{deser-action}, is said to be conformally coupled to gravity.\footnote{In Ref. \cite{padilla} the authors investigated the most general actions containing a single scalar field and two scalar fields coupled to gravity that lead to second order field equations in four dimensions (4D) and posses local scale invariance.} Hence, for instance, the following action \cite{prester, bars, bars_1, bars_2, carrasco, quiros_1, jackiw, alpha, alpha_2, alpha_3, alpha_4}:

\bea S=\int d^4x\sqrt{|g|}\left[\frac{\left(\phi^2-\sigma^2\right)}{12}\,R+\frac{1}{2}(\der\phi)^2-\frac{1}{2}(\der\sigma)^2\right],\label{bars-action}\eea is also invariant under \eqref{scale-t} provided that the additional scalar field $\sigma$ transforms in the same way as $\phi$: $\sigma\rightarrow\Omega\,\sigma$, since in this case both $\phi$ and $\sigma$ are conformally coupled to gravity. The fact that in \eqref{deser-action} the kinetic energy term for the gauge field enters with the wrong sign is not problematic since $\phi$ may be gauged away without physical consequences. The same argument applies to \eqref{bars-action} where, in order for the coupling $\propto (\phi^2-\sigma^2)^{-1}$ to be positive, the scalar $\phi$ must have a wrong sign kinetic energy -- just like in \eqref{deser-action} -- potentially making it a ghost. However, the local Weyl gauge symmetry compensates, thus ensuring the theory is unitary \cite{bars, bars_1, bars_2}.  


Before going into detail about the aim and scope of the present investigation it is necessary to make a step aside and to give certain specifications on the kind of conformal transformations we shall consider in this paper. The Weyl rescalings are composed of a conformal transformation of the metric: 

\bea g_{\mu\nu}\rightarrow\Omega^{-2}g_{\mu\nu},\label{conf-t}\eea plus simultaneous rescalings of the other fields $\Phi_i$ in the theory,

\bea \Phi_i\rightarrow\Omega^w\Phi_i,\label{conf-t-fields}\eea according to their conformal weight $w$. As discussed in \cite{fulton_rmp_1962} conformal transformations can be formulated in different ways so that it is very important to distinguish these different formulations because they have different physical interpretation and because equations belonging to different formulations may easily be confused. Such a confusion can lead to mathematical inconsistencies even when two formulations are equivalent. Accordingly, in the present paper in order to be specific, we assume that the conformal transformation of the metric within the Weyl rescalings does not represent a diffeomorphism or, properly, a conformal isometry (see \cite{dicke-1962} or the appendix D of Ref. \cite{wald_book_1984}). Moreover, the spacetime points -- same as spacetime coincidences or events -- as well as the spacetime coordinates that label the points in spacetime, are not modified or altered by the conformal transformations in any way. 


It is common wisdom that the universe is not Weyl invariant since there are masses and various scales that break the Weyl symmetry \cite{deser}. A typical argument goes like this. Consider the prototype action \eqref{deser-action} without dimensionful parameters. In this Weyl invariant gravity theory only traceless matter (massless fields), can be consistently coupled to gravity. Hence, in order to allow for the coupling of other matter fields to gravity, as well as for the arising of dimensionful parameters, it is necessary to break the Weyl symmetry. One of the simplest symmetry breaking terms that can be added to the action \eqref{deser-action} reads \cite{deser}: 

\bea \frac{1}{2}\int d^4x\sqrt{|g|}\,m^2\phi^2,\label{symm-break}\eea where the mass $m$ of the scalar field is assumed unchanged by the Weyl rescalings.\footnote{In the bibliography there are found other transformation laws for mass parameters. In the references \cite{fulton_rmp_1962, dicke-1962} (see also \cite{faraoni_prd_2007}), for instance, any mass parameter $m$ is transformed under the Weyl rescalings \eqref{scale-t} according to: $m\rightarrow\Omega\,m$. Under this assumption the quantity $\sqrt{|g|}\,m^2\phi^2$ is preserved by the Weyl transformation so that the mass term \eqref{symm-break} does not break the Weyl symmetry.} Other possibilities to break Weyl symmetry include a kind of a symmetry-breaking potential $\propto(\phi^2-v^2)^2$, where $v$ is a mass parameter \cite{zee_prd_1981}, or a symmetry-breaking Lagrangian involving the pseudo-Riemannian curvature scalar $R$ and the mass of a self-interacting scalar field as in \cite{tann}. A different approach is followed in \cite{bars_1} where the breakdown of Weyl symmetry is realized by gauge fixing the value of the scalar field to a constant for all spacetime.


Is there any other way around than Weyl symmetry breaking in order to allow for a consistent matter coupling to gravity? In \cite{waldron, waldron-1} it has been argued that Weyl invariance should be viewed in the same manner as general coordinate invariance: all theories should respect Weyl invariance. In these references, through Weyl invariant coupling of scale and matter fields, the authors construct theories that unify massless, massive, and partially massless excitations. The method used by the authors relies on tractor calculus -- mathematical machinery allowing Weyl invariance to be kept manifest at all stages.

The aim of this paper is a bit more modest. Here we want to approach Weyl invariance from the geometrical and physical standpoints in a rather straightforward way, without the need for a special mathematical machinery as in \cite{waldron, waldron-1} and without appealing to any new theoretical framework: We shall explore the physical consequences of a version of a known Weyl invariant theory of gravity called here as conformal general relativity. Several outstanding mathematical features of similar theories have been explored before \cite{scholz, romero_ijmpa, romero_cqg, lobo, romero_et_all, pucheu}. We shall discuss, in particular, on the physical implications of gauge freedom -- an inevitable consequence of Weyl invariance. 

The main idea which the CGR is based on is to replace the pseudo-Riemannian geometrical structure of background spacetimes by Weyl integrable geometry (see the appendix \ref{sect-app-a} and references therein). The present version of CGR is based, besides, on the following specific postulate: Only the fundamental fields are transformed by the scale transformations. Hence, dimensionless as well as dimensionful constants, including the fundamental constants of nature such as the Planck constant $\hbar$, the speed of light, the Planck mass, etc., are not transformed by the Weyl rescalings. This postulate makes it possible to have Weyl invariant laws of gravity that are compatible with the existence of certain scales such as the Planck and the EW mass scales. This theory contains general relativity as a particular gauge. 


The plan of the paper is as follows. The most important features of the present version of CGR, including the consequences of the gauge freedom, are explained in section \ref{sect-egr}. In section \ref{sect-redshift} we discuss on the WIG geodesics and the way the redshift arises in Weyl-integrable spaces. The issue about the physically meaningful quantities of this gravitational theory is exposed in section \ref{sect-gauge-inv}, where in subsection \ref{subsect-sing} the singularity issue is investigated on the basis of the analysis of the Weyl gauge invariants of the theory. A discussion on the measured Newton's constant is provided in section \ref{sect-newton-c}, while in section \ref{sect-weyl-smp} it is shown how to consistently couple the standard model of particles, in such a way as to preserve the Weyl symmetry of the theory even after the breakdown of EW symmetry. This happens to be an important requirement of WIG-based theories of gravity since, otherwise, point-dependent mass units are not allowed. In order to illustrate with concrete examples the implications of gauge freedom, as well as of physical equivalence of the different gauges, several physically interesting cosmological solutions of CGR are explored in section \ref{sect-cosmo}. It is demonstrated in section \ref{sect-infl} that, due to gauge freedom, the flatness, horizon and relict particle abundances, among other well-known puzzles of the hot bigbang scenario, do not arise in the framework of the CGR cosmological models. Another outstanding consequence of gauge freedom in our setup is its ability to overcome the mass hierarchy and the cosmological constant problems. In section \ref{sect-puzzles} it is shown how these issues are settled in the proposed version of CGR. The results of this paper are discussed in section \ref{sect-discuss} where we compare the present setup with other similar approaches, while conclusions are given in section \ref{sec-conclu}. 

For completeness of the exposition we have included an appendix section. In the appendix \ref{sect-app-a} a brief account of Weyl integrable geometry is given, while in the appendix \ref{sect-app-b} we give the details of the derivation of the motion equations of our setup. A detailed discussion on the role of the choice of the transformation properties of the fields and of the constants of the theory under Weyl rescalings is provided in the appendix \ref{sect-app-d}. The latter issue requires of a careful discussion since, depending on the specific postulates of the theory, it may happen that dimensionless fields may be transformed by the Weyl rescalings while dimensionful constants like the Planck constant $\hbar$, the electric charge of the electron $e$, and even mass parameters like in \eqref{symm-break}, are kept unchanged by these transformations. Here we use the mostly positive signature of the metric: $(-,+,+,+)$ and the following units' convention: $\hbar=c=1$, is assumed.


\section{conformal general relativity}\label{sect-egr}

An aspect of local scale invariance of gravity theories that is not usually discussed, is related with its geometrical implications: Invariance under Weyl rescalings \eqref{scale-t} is meaningless until a geometrical background is specified, where by geometrical background we do not understand just a metric but a whole geometrical setup, i. e., a set of geometrical laws that define a geometrical structure such as, for instance, Riemann geometry, Weyl geometry, etc. Conformal general relativity is a modification of general relativity where the Riemannian affine structure of the background spacetimes is replaced by a Weyl-integrable structure \cite{novello, scholz_1, scholz_2, quiros_prd_2000, quiros_npb_2002, scholz, romero_ijmpa, romero_cqg, lobo, romero_et_all, pucheu}. Here we shall explore the geometrical and physical consequences of a version of CGR that is based on the following postulates:

\begin{itemize}

\item The affine structure of the spacetime is determined by Weyl-integrable geometry and not by the laws of Riemann geometry.

\item Only the fields (this includes the fundamental fields like the metric, the gauge scalar, the Higgs, the spinors and other matter fields) are transformed by the Weyl rescalings.\footnote{It is understood that the ratio of fields with the same conformal weight, being in principle point dependent quantities, is not transformed by the conformal transformations.} The dimensionless as well as the dimensionful constant parameters, including the fundamental constants of nature such as the Planck constant $\hbar$, the speed of light, the Planck mass, the fine structure constant, $\alpha$, etc., are not transformed by the Weyl rescalings.

\end{itemize} 

Sometimes we shall call the actual (dimensionless and dimensionful) constants as ``bare constants'' in contrast to ``point-dependent constants'' which are obtained as the product of a bare constant with an appropriate power of the gauge scalar. The fields in the theory transform according to their conformal weight under the Weyl rescalings:

\bea g_{\mu\nu}\rightarrow\Omega^{-2}g_{\mu\nu},\;\;\vphi\rightarrow\vphi+2\ln\Omega.\label{weyl-t}\eea Their transformation properties can be summarized as follows. The scalars like in \eqref{deser-action} are transformed like, $\phi\rightarrow\Omega\,\phi$, while dilaton-like scalars, such as the gauge scalar, transform like, $\vphi\rightarrow\vphi+2\ln\Omega$. For the vector-potentials we have that, $A_a^\mu\rightarrow\Omega^2\,A_a^\mu$ ($A^a_\mu\rightarrow\,A^a_\mu$) and similar for the 4-momentum, $p^\mu$, which appears in gauge invariant combination with the vector potential: $p_\mu+e A_\mu$ ($e$ is the electric charge of the electron). Meanwhile, for the fermion fields: $\psi\rightarrow\Omega^{3/2}\,\psi$, etc. Let us mention that there are vectors like the 4-velocity: $u^\mu=dx^\mu/ids$, which under the Weyl recalings do not transform in the same way as $p^\mu$ and $A^\mu$: $u^\mu\rightarrow\Omega\,u^\mu$, $u_\mu\rightarrow\Omega^{-1}u_\mu.$ This is compensated by the transformation of the mass $m\rightarrow\Omega\,m$, so that the required transformation of the four momentum is achieved: $$p^\mu=m\,u^\mu\rightarrow\Omega^2\,p^\mu\;\;(p_\mu\rightarrow p_\mu).$$ In this regard it has to be said that $p_\mu$ -- as well as $p_\mu+e A_\mu$ -- is a Weyl gauge vector, while $u_\mu$ is not.


\subsection{Equations of motion}

The mathematical foundation of the present theory is given by the following Weyl invariant action:

\bea S=\int d^4x\sqrt{|g|}\left[\frac{M^2_\text{Pl}(\vphi)}{2} R^{(w)}-\lambda\,e^{2\vphi}+{\cal L}_m\right].\label{winv-action}\eea More complex Weyl invariant Lagrangians may include higher-order curvature terms like: $${\cal L}_\text{high}=\left[aR^2_{(w)}+b R^{(w)}_{\mu\nu}R_{(w)}^{\mu\nu}+c e^{-\vphi}R^{(w)}_{\mu\nu\tau\lambda}R_{(w)}^{\mu\nu\tau\lambda}\right],$$ where $a$, $b$ and $c$ are constants. However, such kind of higher-order terms usually bring with them additional degrees of freedom -- including ghosts -- and we shall not include them in the present study.

From the action \eqref{winv-action} the following motion equations can be derived:

\bea &&G^{(w)}_{\mu\nu}=\frac{1}{M^2_\text{Pl}(\vphi)}\left[T^{(m)}_{\mu\nu}-\lambda\,e^{2\vphi}g_{\mu\nu}\right],\nonumber\\
&&M^2_\text{Pl}(\vphi)R^{(w)}=4\lambda\,e^{2\vphi}-T_{(m)}\;\Leftrightarrow\;\Box\vphi+\frac{1}{2}\left(\der\vphi\right)^2-\frac{R}{3}=\frac{e^{-\vphi}}{3M^2_0}\,T_{(m)}-\frac{4\lambda\,e^\vphi}{3M^2_0},\label{winv-feq}\eea where $M^2_\text{Pl}(\vphi)=M^2_0\,e^\vphi$ is the point-dependent Planck mass, $M^2_0$ is the bare Planck mass squared and $\lambda$ is also a bare free constant, $R$ is the standard (Riemannian) curvature scalar, and $T_{(m)}=g^{\mu\nu}T^{(m)}_{\mu\nu}$, is the trace of the standard stress-energy tensor (SET). In the above equations the sub(supra)index `$(w)$' means that the given quantity and/or geometrical object is defined in terms of the WIG affine connection \eqref{weyl-aff-c} and the associated operators.

Since deriving the motion equations \eqref{winv-feq} is a rather non-trivial task, in the appendix \ref{sect-app-b} we expose the details of the derivation. In every step during the derivation process it is mandatory to keep in mind that in WIG spaces (see appendix \ref{sect-app-a}) we must require fulfillment of the metric compatibility condition \eqref{met-cond}: $$\nabla^{(w)}_\lambda g_{\mu\nu}=-\der_\lambda\vphi\,g_{\mu\nu}\;\Leftrightarrow\;\nabla^{(w)}_\lambda g^{\mu\nu}=\der_\lambda\vphi\,g^{\mu\nu},$$ so that, for instance: $$\nabla^{(w)}_\lambda\sqrt{|g|}=-\frac{1}{2}\sqrt{|g|} g_{\mu\nu}\nabla^{(w)}_\lambda g^{\mu\nu}=-2\sqrt{|g|}\,\der_\lambda\vphi.$$ As a consequence variations with respect to the gauge field are not independent from variations with respect to the metric, since variations with respect to $\vphi$ induce variations of the metric according to: 

\bea \delta g_{\mu\nu}=-\delta\vphi\,g_{\mu\nu}\;\Leftrightarrow\;\delta g^{\mu\nu}=\delta\vphi\,g^{\mu\nu}.\label{delta-g}\eea This outstanding feature of WIG spaces is to be taken into account in particular while deriving the motion equation for the gauge field $\vphi$ (equation in the second line in \eqref{winv-feq}).


Notice that there is not an independent motion equation for the gauge field $\vphi$ since the equation in the second line in \eqref{winv-feq} coincides with the trace of the WIG-Einstein's equation, i. e., of the equation in the first line in \eqref{winv-feq}. Hence, the $\vphi$-field is not dynamical: it may be gauged away without physical consequences. This is an inevitable consequence of gauge freedom in connection with Weyl invariance. Several mathematical aspects of similar setups have been formerly explored in \cite{scholz, romero_ijmpa, romero_cqg}. We shall comment on these and other related developments in section \ref{sect-discuss}, in order to make clear the differences with our setup.


\subsection{Weyl invariant stress-energy tensor}

The contracted Bianchi identity in the WIG reads: $\nabla^\mu_{(w)}G^{(w)}_{\mu\nu}=0$, while the Weyl-integrable metricity condition is given by (see appendix \ref{sect-app-a}): $\nabla^\sigma_{(w)}g_{\mu\nu}=-\der^\sigma\vphi g_{\mu\nu}$. In consequence the conservation equation that follows from the Einstein's equation \eqref{winv-feq} can be written as:

\bea \nabla^\mu_{(w)}T^{(m)}_{\mu\nu}=\der^\mu\vphi\,T^{(m)}_{\mu\nu},\label{conserv-eq}\eea where the RHS is not a source term since it is originated from the fact that the units of measure of the stresses are point-dependent quantities in WIG. 

It may be convenient to introduce the WIG-SET for the matter: 

\bea T^{(w,m)}_{\mu\nu}:=e^{-\vphi}T^{(m)}_{\mu\nu},\label{wig-set}\eea so that the conservation equation \eqref{conserv-eq} takes the ``formal'' conservation equation look: 

\bea \nabla^\mu_{(w)}T^{(w,m)}_{\mu\nu}=0,\label{wig-conserv-eq}\eea while the WIG-Einstein's equations of motion read,

\bea G^{(w)}_{\mu\nu}=\frac{1}{M^2_0}\left[T^{(w,m)}_{\mu\nu}-\lambda\,e^\vphi g_{\mu\nu}\right].\label{wig-feq}\eea 

The above is not only a convenient definition of the stress-energy tensor, but it is the one carrying the gauge-invariant meaning. Notice that the WIG-SET, $T^{(w,m)}_{\mu\nu}$, shares the same transformation properties of the WIG Einstein's tensor $G^{(w)}_{\mu\nu}$ under the Weyl rescalings \eqref{weyl-t}. As we shall see in section \ref{sect-newton-c}, the fact that it is $T^{(w,m)}_{\mu\nu}$ and not the standard matter SET, $T^{(m)}_{\mu\nu}$, the one that has the physical meaning, entails that the measured Newton's constant is the bare constant, $G_N=1/8\pi M_0^2$, and not the related point-dependent gravitational coupling, $G_N(\vphi)=1/8\pi M^2_\text{Pl}(\vphi)$.

It is important to notice that the fundamental dimensionful bare constants such as the Plack mass, $M_0$, are included in the action \eqref{winv-action} from the start. In this regard, the theory \eqref{winv-action} is to be considered as less fundamental than, for instance, the ones in Refs. \cite{smolin_npb_1979, zee_prd_1981, cheng_prl_1988, prester, bars, bars_1, bars_2, carrasco, jackiw, alpha, alpha_2, alpha_3, alpha_4} where the fundamental constants arise as consequence of the break down of conformal symmetry. It is clear that the emergence of fundamental scales can not be addressed within this setup. Besides, the action \eqref{winv-action} differs from the one in \cite{prester, bars, bars_1, bars_2, carrasco, jackiw, alpha, alpha_2, alpha_3, alpha_4}, in that the underlying geometric structure is WIG. Since in this case not only the orientation, but also the length of vectors are point-dependent quantities (see the appendix \ref{sect-app-a}), this entails that the units of measure themselves are point dependent units as well.


\subsection{Gauge freedom: the GR gauge}\label{subsect-g-free}

It is apparent from \eqref{winv-feq} that there is not an independent equation of motion for the gauge field $\vphi$.  The WIG-Einstein's equation \eqref{wig-feq} and the conservation equation \eqref{wig-conserv-eq}, are the only independent motion equations of the above Weyl invariant theory of gravity. This property is consequence of invariance of the theory under the Weyl rescalings \eqref{weyl-t}. Hence, in addition to the four degrees of freedom to make diffeomorphisms, there is an additional degree of freedom to make conformal transformations. In other words: we are free to choose any function $\vphi=\vphi(x)$ we want, provided it is continuous and (at least twice) differentiable.\footnote{As a matter of fact one may choose at will either the gauge scalar, $\vphi$, or one of the components of the metric tensor.}

A particularly simple and outstanding gauge is the one given by the choice: $\vphi=\vphi_0$, where $\vphi_0$ is a constant. Under this choice the affine connection of WIG spaces is transformed into the Christoffel symbols of the metric and the metricity condition of WIG is transformed into the Riemann metricity property: the metric tensor is covariantly constant. This means, in turn, that the Weyl-integrable spaces are transformed into spaces with (pseudo)Riemann geometric structure, as those in GR. In correspondence with this choice of gauge, the motion equation \eqref{wig-feq} transforms into the Einstein's equation of general relativity: $$G_{\mu\nu}=8\pi G_N\left(T^{(m)}_{\mu\nu}-\rho_\lambda g_{\mu\nu}\right),$$ with $$8\pi G_N\equiv\frac{1}{M^2_0},\;\;\rho_\lambda\equiv\lambda\,e^{\vphi_0}.$$ The above motion equation is supplemented with the standard conservation equation: $\nabla^\mu T^{(m)}_{\mu\nu}=0$. In other words, the choice $\vphi=\vphi_0$ corresponds to plain general relativity, and we call this as the GR gauge. Hence, GR is one of the infinitely many physically equivalent representations of the Weyl invariant laws of gravity \eqref{winv-action}. This is why the above Weyl invariant theory of gravity is called as conformal general relativity. 

We stress that, assuming that $\vphi$ is a smooth function of the spacetime point, means that a small neighborhood of every point $P$ in the spacetime can be always found where $\vphi=\vphi_0$ is a constant and the GR gauge is singled out. This intuitive fact entails that, locally, CGR is well approached by GR as required by local Lorentz invariance of the laws of nature.


\section{On WIG geodesics and the redshift}\label{sect-redshift}

One of the indisputable assumptions of the conformal transformations procedure is that the co-variant components of the vector potential $A_\mu$ are not transformed by the conformal transformation in \eqref{weyl-t}. Hence, consistency of the procedure requires that the co-variant components of any other Weyl-gauge vector, be it the 4-wavevector $k_\mu$ of a photon, or the 4-momentum of a particle $p_\mu$, would not be transformed by the Weyl rescalings: $(A_\mu,k_\mu,p_\mu,...)\rightarrow(A_\mu,k_\mu,p_\mu,...)$. This means that under the aforementioned transformations, $$(A^\mu,k^\mu,p^\mu,...)\rightarrow\Omega^2(A^\mu,k^\mu,p^\mu,...).$$ Hence, the Weyl invariant WIG geodesic equations for particles with point dependent mass $m(\vphi)=m_0 e^{\vphi/2}$ and 4-momentum, $$p^\mu=m(\vphi)\frac{dx^\mu}{ids},$$ read (compare with equation \eqref{weyl-geod-1}): 

\bea \frac{dp^\mu}{ds}+\Gamma^\mu_{\sigma\lambda}\frac{dx^\sigma}{ds}p^\lambda-\der_\lambda\vphi\frac{dx^\lambda}{ds}p^\mu=0.\label{4-p-geod}\eea The 4-wavevector $k^\mu=(w,{\bf k})$ of a photon obeys the same geodesic equation after simultaneously replacing $p^\mu\rightarrow k^\mu$, and the line element by the differential of an affine parameter $\sigma$ along the photon path: $ds\rightarrow d\sigma$. We recall that, under the Weyl rescalings \eqref{weyl-t}, $(ds,d\sigma)\rightarrow\Omega^{-1}(ds,d\sigma)$. The subtle point here is that, if in the above geodesic equations replace the WIG affine connection, $\Gamma^\mu_{\sigma\lambda}$, by its expression through the Christoffel symbols of the metric, $\{^\mu_{\sigma\lambda}\}$ in \eqref{weyl-aff-c}, for the geodesic of a particle with 4-momentum we get: $$\frac{dp^\mu}{ds}+\{^\mu_{\sigma\lambda}\}\frac{dx^\sigma}{ds}p^\lambda-\frac{1}{2}g_{\sigma\lambda}\frac{dx^\sigma}{ds}p^\lambda\der^\mu\vphi=0,$$ while for a photon, since $g_{\sigma\lambda}dx^\sigma dx^\lambda=0$, the geodesic: 

\bea \frac{dk^\mu}{d\sigma}+\{^\mu_{\sigma\lambda}\}\frac{dx^\sigma}{d\sigma}k^\lambda=0,\label{photon-geod}\eea coincides with the geodesic of a photon in a Riemannian spacetime. This is another way to show that the geodesics of massless particles are not affected by the conformal transformations of the metric. In other words, the photon is blind to the affinity of the spacetime: it does not differentiate a Riemannian space from a Weyl-integrable one. Yet, the overwhelming amount of information about our universe is obtained through observations involving the light.

Consider next a static FRW space (the scale factor is a constant), so that the Christoffel symbols of the metric all vanish. From the geodesic equation for particles with the mass it follows that the 4-momentum changes along the geodesic, meanwhile, from the photon's geodesic equation it follows that the 4-wavevector is unchanged under parallel transport along the null-geodesic. Let us assume that the (rest) energy of a given atomic transition in a given spacetime point $P$, is given by: $\Delta m(\vphi(P))=\Delta m_0\exp(\vphi(P)/2)$. As a result of the transition a photon of energy $\hbar\omega_P=\Delta m(\vphi(P))$ is emitted. Suppose now that an identical atom that is placed at another spacetime point $P_0$, also emits a photon as a result of the same transition. The energy of the emitted photon is given by: $\hbar\omega_{P_0}=\Delta m_0\exp(\vphi(P_0)/2).$ Let us assume that the photon emitted at $P$ reaches $P_0$ so that its energy can be compared with the energy of the photon emitted at $P_0$. Since, according to \eqref{photon-geod}, in a static universe the energy carried by the photon is unchanged along the photon's path, the relative magnitude of the redshift in the case discussed here, is given by: $$\frac{\Delta\omega}{\omega_P}=\frac{\left(\omega_{P_0}-\omega_P\right)}{\omega_{P_0}}=1-\exp\left(\frac{\vphi(P)-\vphi(P_0)}{2}\right).$$ Hence, the occurrence of a non-vanishing redshift in a static WIG universe is due to the fact that the masses of particles -- including atoms, etc. -- that emit photons, are point-dependent quantities, while the energy of the photons themselves is unchanged along photons path's in a flat spacetime background.


\section{Weyl gauge invariants}\label{sect-gauge-inv}

In the CGR the quantities with the physical meaning are necessarily invariant under the Weyl rescalings \eqref{weyl-t}. Actually, even if in the experiments one is not always able to measure directly the invariants, these are the ones that preserve the physical meaning independent of the coordinates and of the gauge chosen. For instance, the gauge invariant measure of the curvature scalar is defined by, $R_*:=e^{-\vphi}R^{(w)}$, while the gauge-invariant measure of spacetime separations can be, $ds^2_*:=e^\vphi ds^2$. Other Weyl invariant quantities are: $$e^{-2\vphi}R^{(w)}_{\mu\nu}R_{(w)}^{\mu\nu},\;e^{-3\vphi}R^{(w)}_{\mu\nu\tau\lambda}R_{(w)}^{\mu\nu\tau\lambda},$$ among others. 

Let us focus, for illustration, in the Weyl invariant Kretschmann scalar: $$K_*:=e^{-3\vphi}R^{(w)}_{\mu\nu\tau\lambda}R_{(w)}^{\mu\nu\tau\lambda}.$$ In the GR gauge, for instance, $$K_*=e^{-3\vphi_0}R_{\mu\nu\tau\lambda}R^{\mu\nu\tau\lambda},$$ where $R_{\mu\nu\tau\lambda}$ is the standard Riemann-Christoffel curvature tensor. Hence, we have that: 

\bea K_\text{GR}=R_{\mu\nu\tau\lambda}R^{\mu\nu\tau\lambda}=e^{-3(\vphi-\vphi_0)}R^{(w)}_{\mu\nu\tau\lambda}R_{(w)}^{\mu\nu\tau\lambda},\label{gr-kretschmann}\eea where $K_\text{GR}$ is the standard (Riemannian) Kretschmann invariant. The above equation will be useful below when we discuss on the singularity issue.

Notice that Riemannian (coordinate) invariant combinations like the Gauss-Bonnet topological term (in 4D): $${\cal L}_\text{GB}=R^2-4R_{\mu\nu}R^{\mu\nu}+R_{\mu\nu\tau\lambda}R^{\mu\nu\tau\lambda},$$ do not always have a WIG equivalent. Within WIG spaces, the following combination can be found instead: 

\bea {\cal L}=R_{(w)}^2-4R^{(w)}_{\mu\nu}R_{(w)}^{\mu\nu}+e^{-\vphi}R^{(w)}_{\mu\nu\tau\lambda}R_{(w)}^{\mu\nu\tau\lambda},\label{wig-gb}\eea which can not be called properly as ``WIG Gauss-Bonnet term'' due to the factor $e^{-\vphi}$ in the third term. This means that it is not actually a topological (boundary) term in 4D since its inclusion in the gravitational action does actually contribute towards the equations of motion.


\subsection{Conformal general relativity and the singularity issue}\label{subsect-sing}

The apparent paradox that arises when comparing solutions with a spacetime singularity in one frame with the same solutions in a conformal frame where the singularity may be absent,\footnote{It has been shown that Weyl invariant dilaton gravity provides a description of black holes without classical spacetime singularities \cite{prester}. The singularities appear due to ill-behavior of gauge fixing conditions, one example being the gauge in which the theory is classically equivalent to GR.} has been discussed several times without a clear resolution \cite{faraoni_prd_2007, quiros_grg_2013, kaloper_prd_1998, quiros_prd_2000}. In Ref. \cite{quiros_prd_2000}, for instance, the issue was discussed under the assumption that the underlying geometrical structure of the background spacetime is WIG, but the action considered is itself not conformal invariant, so that the conformal frames are not actually physically equivalent. 

Here we shall re-visit the issue by approaching to it from the point of view of the CGR. In this theory the different conformal frames or gauges, are actually physically equivalent given that, not only the action of the theory \eqref{winv-action} and the derived motion equations \eqref{wig-feq} are invariant under Weyl rescalings \eqref{weyl-t}, but, also the assumed WIG-structure of the background spacetimes warrants that the geodesics of the geometry are conformal invariant.  

Since the quantities that have the physical meaning are those which are not only invariant under general coordinate transformations but, at the same time, are also invariant under the Weyl rescalings \eqref{weyl-t} -- here we call these as Weyl gauge (or just gauge) invariant quantities -- our discussion will rely exclusively on the gauge invariants. Take, for instance, the gauge-invariant measure of spacetime separations: $ds_*^2=e^\vphi g_{\mu\nu}dx^\mu dx^\nu$. Consider the GR-gauge, where $\vphi=\vphi_0$ (for simplicity let us take $\vphi_0=0$). We have that, $ds_*^2=ds^2_\text{GR}=g^\text{GR}_{\mu\nu}dx^\mu dx^\nu.$ This entails the following relationship between the GR metric tensor and a given conformal metric:

\bea g_{\mu\nu}=e^{-\vphi}g_{\mu\nu}^\text{GR}.\label{metric-rel}\eea Let us further assume, for definiteness, the spherically symmetric Schwarzschild GR vacuum metric, $$ds^2_\text{GR}=-\left(1-\frac{2m}{r}\right)dt^2+\left(1-\frac{2m}{r}\right)^{-1}dr^2+r^2d\Omega^2,$$ where $d\Omega^2\equiv d\theta^2+\sin^2\theta d\phi^2$. Since we can freely choose the gauge scalar, let us set:

\bea \vphi(r)=2q\ln\left(1-\frac{2m}{r}\right),\label{vphi-r}\eea where $q$ is an arbitrary constant that parametrizes our choice for the gauge scalar. Hence, we have a set of physically equivalent WIG descriptions of the laws of gravity \eqref{winv-action}, \eqref{winv-feq} given by: $\{({\cal M}_{(q)},g^{(q)}_{\mu\nu},\vphi_{(q)}):\,q\geq 0\}$, where ${\cal M}_{(q)}$ represents the $q$-th spacetime manifold.\footnote{The values $q<0$ are not considered since, as shown in Ref. \cite{quiros_prd_2000}, in this case one gets a set of spacetimes with naked singularities.} Notice that general relativity is included in the above equivalence class as the representation specified by the choice, $q=0$. For the $q$-th representation the metric reads:

\bea ds^2=-\left(1-\frac{2m}{r}\right)^{1-2q}dt^2+\left(1-\frac{2m}{r}\right)^{-1-2q}dr^2+\rho^2d\Omega^2,\label{ds-q}\eea where, $$\rho(r)=\frac{r}{(1-2m/r)^q},$$ is the proper radial coordinate. Since $\rho$ could be non-negative, then: $2m\leq r<\infty$. Besides, the proper radial coordinate is a minimum at $r_\text{min}=2(1+q)m$, where 

\bea \rho_\text{min}=\frac{(1+q)^{1+q}}{q^q}\,2m.\label{min-rho}\eea 

In order to discuss on the occurrence of spacetime singularities it is useful to write the relationship between the GR Kretschmann scalar, $K_\text{GR}=R_{\mu\nu\tau\lambda}R^{\mu\nu\tau\lambda}=48m^2/r^6$, and the one for the conformal WIG representation, $K_{(w)}=R^{(w)}_{\mu\nu\tau\lambda}R_{(w)}^{\mu\nu\tau\lambda}$, given in \eqref{gr-kretschmann} with $\vphi_0=0$: 

\bea K_{(w)}=e^{3\vphi}K_\text{GR}=\frac{48m^2}{r^6}\left(1-\frac{2m}{r}\right)^{6q}=\frac{48m^2}{\rho^6}.\label{kretsh-rel}\eea 

A class of spacetimes with two asymptotically flat spatial infinities is obtained: one at $r\rightarrow\infty$ and the other one at $r\rightarrow 2m$ where, in both cases, $\rho\rightarrow\infty$ while $K_{(w)}\rightarrow 0$. These spatial infinities are joined by a throat with minimum value of the proper radial coordinate, $\rho_\text{min}$, given by \eqref{min-rho}, where the curvature is a maximum: $$K_{(w)}^\text{max}=\frac{3q^{6q}}{4(1+q)^{6(1+q)}m^4}.$$ A similar result was discussed in Ref. \cite{quiros_prd_2000}, however, as already mentioned, in that reference the BD theory was considered so that the analysis performed was not based on the gauge invariants like in the present section. Other discussions on the singularity issue can be found \cite{faraoni_prd_2007, kaloper_prd_1998} where, once again, due to the lack of invariance under the Weyl rescalings \eqref{weyl-t}, the analysis could not be based on the gauge invariants.\footnote{It has been shown in Ref. \cite{veermae} that the existence of wormhole vacua in conformal (Weyl invariant) theories of gravity is a completely general and unavoidable feature of conformal gravity.}

The result we have just obtained is a clear example of the fact that a given spacetime singularity existing in one or in several physically equivalent frames, may be safely avoided in other physically equivalent representations. In the present case, a Schwarzschild black hole with a singularity at $r=0$, that is enclosed by an event horizon at $r=2m$ in the GR gauge, is replaced by wormhole spacetimes which are free of singularities, in a class of conformal (physically equivalent) gauges: $\{({\cal M}_{(q)},g^{(q)}_{\mu\nu},\vphi_{(q)}):\,q\geq 0\}$.


\section{Measured Newton's constant}\label{sect-newton-c}

As already stressed the physically meaningful SET in the WIG-Einstein equations \eqref{wig-feq} is the gauge invariant matter SET, $T^{(w,m)}_{\mu\nu}$ (\ref{wig-set}), which is the one conserved in WIG spacetimes and the one that shares the same transformation properties that the WIG Einstein's tensor $G^{(w)}_{\mu\nu}$. Take as an illustration a perfect fluid with stress-energy tensor

\bea T^{(w,m)}_{\mu\nu}=\left[\rho^{(w)}_m+P^{(w)}_m\right]u_\mu u_\nu+P^{(w)}_m g_{\mu\nu},\label{perf-fluid-set}\eea where $u^\mu=dx^\mu/d\tau$ ($d\tau=-ids$), and

\bea \rho^{(w)}_m=e^{-\vphi}\,\rho_m,\;P^{(w)}_m=e^{-\vphi} P_m,\,P^{(w)}_m=(\gamma-1)\rho^{(w)}_m,\label{baro}\eea are the WIG energy density and barotropic pressure respectively ($\gamma$ is the barotropic index). The energy density measured by physical observers which are co-moving with the perfect fluid, i. e., their 4-velocity $u^\mu$ coincides with that of the fluid itself, is indeed

\bea \rho^{(w)}_m=T^{(w,m)}_{\mu\nu}u^\mu u^\nu,\label{wig-rho}\eea and not $\rho$.\footnote{Notice that under \eqref{weyl-t} the standard (Riemannian) GR energy density and the one which is measured by WIG observers transform differently: $$\rho_m\rightarrow\Omega^4\rho_m,\;\rho^{(w)}_m\rightarrow\Omega^2\rho^{(w)}_m.$$} As a consequence the gravitational coupling measured in Cavendish-type experiments -- see the related computations below -- is just the usual Newton's constant $G_N$, a fact which is evident also from the WIG-Einstein's equations \eqref{wig-feq}.


\subsection{Newtonian limit}\label{subsect-weak-f}

In the Newtonian limit where particles move slowly $dx^i/ds\ll dt/ds$, and only weak static gravitational field is considered, assuming small departure from Riemannian geometry $\der_i\vphi\ll 1$: 

\bea &&g_{\mu\nu}=\eta_{\mu\nu}+h_{\mu\nu},\;\vphi=\vphi_0+\phi,\nonumber\\
&&|h_{\mu\nu}|\ll 1,\;\phi\ll 1,\;\frac{\der h_{\mu\nu}}{\der t}=0,\;\frac{\der\phi}{\der t}=0,\nonumber\eea we have for the geodesic motion: $$\frac{d}{ds}\left(\frac{dx^\mu}{ds}\right)+\Gamma^\mu_{\;00}\left(\frac{dt}{ds}\right)^2=0,$$ so that $$\frac{d^2x^i}{dt^2}=\frac{1}{2}\der_i (h_{00}-\phi)\;\Rightarrow\;h_{00}-\phi=-2\Phi,$$ where $\Phi$ is the Newtonian gravitational potential. In the same limit, since $$T^{(w,\text{m})}_{00}=\rho^{(w)}\gg|T^{(w,\text{m})}_{ik}|,\;T^{(w,\text{m})}=-T^{(w,\text{m})}_{00},$$ the WIG-Einstein field equations lead to ($\nabla^2\equiv \der^2_i$)

\bea R^{(w)}_{00}=\frac{1}{2M^2_0}\,T^{(w,\text{m})}_{00}\;\Rightarrow\;\nabla^2(h_{00}-\phi)=-\frac{\rho^{(w)}}{M^2_0}\;\Rightarrow\;\nabla^2\Phi=\frac{\rho^{(w)}}{2M^2_0}=4\pi G_N\rho^{(w)},\nonumber\eea so that, $$\Phi(x)=-G_N\int_{\Omega_\textsc{m}}d^3x'\frac{\rho^{(w)}({\bf x}')}{|{\bf x}-{\bf x}'|},$$ where $\Omega_\textsc{m}$ is the volume occupied by the mass distribution. For a mass point at the origin we get: $\Phi=-G_N\,m/r$, where $m$ is the mass of the point source and $r=|{\bf x}|$. 

From the above discussion it is apparent that, given that the WIG-gauge SET \eqref{wig-set} (see also \eqref{perf-fluid-set}) in the WIG-Einstein equation \eqref{wig-feq} is the one carrying the physical meaning, it is the bare Newton's constant $G_N=M^{-2}_0/8\pi$, and not the point-dependent coupling, $G(\vphi)=M^{-2}_\text{Pl}(\vphi)/8\pi$, the one measured in Cavendish experiments. In other words, the measured Newton's constant is a genuine gauge constant that is not transformed by the Weyl rescalings and, consequently, does not depend on the spacetime point.


\section{Weyl invariant Standard Model of Particles}\label{sect-weyl-smp}

In order to incorporate the standard model of elementary particles in a Weyl invariant way into the theory \eqref{winv-action}, it suffices to promote the -- otherwise constant -- mass parameter $v_0$ to a point-dependent quantity \cite{scholz, bars}: $v=v_0\exp{(\vphi/2)}$. For other alternatives of coupling the SMP to gravity in a Weyl invariant way see, for instance, Ref. \cite{bars_1, scholz, pawlowski}.

Following our reasoning line the EW Lagrangian for the Higgs field can be written as:

\bea {\cal L}_H=-\frac{1}{2}|DH|^2-\frac{\lambda'}{2}\left[|H|^2-v_0^2e^\vphi\right]^2,\label{higgs-lag}\eea with $|H|^2\equiv H^\dag H$ and $|DH|^2\equiv g^{\mu\nu}(D_\mu H)^\dag(D_\nu H)$, where the gauge covariant derivative: 

\bea D_\mu H\equiv(D^*_\mu-\frac{1}{2}\der_\mu\vphi)H.\label{gauge-der}\eea Here $$D^*_\mu H\equiv\left(\der_\mu+\frac{i}{2}g W_\mu^k\sigma^k+\frac{i}{2}g'B_\mu\right)H,$$ is the standard gauge covariant derivative of the EW theory with $W_\mu^k=(W_\mu^\pm,W_\mu^0)$-- the $SU(2)$ bosons, $B_\mu$-- the $U(1)$ boson, $\sigma^k$-- the Pauli matrices and $(g,g')$-- the gauge couplings. Notice that, under the Weyl rescalings \eqref{weyl-t}: $$|H|\rightarrow\Omega\,|H|,\;\left(W_\mu^k,B_\mu\right)\rightarrow\left(W_\mu^k,B_\mu\right),$$ while the enlargement of the gauge covariant derivative to include the derivative of the Weyl gauge scalar \eqref{gauge-der}, allows the corresponding operator to preserve, also, the Weyl symmetry. This entails that under \eqref{weyl-t}, $D_\mu\rightarrow D_\mu$, and that, besides, the gauge derivative commutes with the conformal factor: $[D_\mu,\Omega]=0$, so that, $$D_\mu H\rightarrow\Omega D_\mu H\;\Rightarrow\;|DH|^2\rightarrow\Omega^4|DH|^2,$$ leading to: ${\cal L}_H\rightarrow\Omega^4{\cal L}_H$, i. e., the EW Higgs field action piece, $S_H=\int d^4x\sqrt{|g|}{\cal L}_H$, is not transformed by the Weyl rescalings. Above we have assumed, in accordance with one of the postulates on which the present setup rests, that the bare constants like $g$, $g'$, $\lambda'$ and $v_0$ are not transformed by \eqref{weyl-t}. 

We want to stress that the only way in which the standard unit of mass can be a point-dependent quantity is that the particles of the SMP acquire point-dependent masses as a result of the symmetry breaking procedure. This is what is achieved in the SMP supported by the Weyl invariant Lagrangian \eqref{higgs-lag}. In this WIG-based model, as a result of ``symmetry breaking,'' the Higgs acquires a point-dependent VEV; $|H|=v_0\,e^{\vphi/2}$, so that the gauge bosons and fermions of the SMP acquire point-dependent masses: $m_p(\vphi)=g_p\,v_0\,e^{\vphi/2}$, where $g_p$ is some gauge coupling. Fermion fields, $\psi$, acquire point-dependent masses through Yukawa interactions of the form: $g_\psi\bar\psi H\psi$ ($g_\psi$ is a Yukawa coupling), where, after EW symmetry breaking the Higgs acquires a point-dependent VEV: $|H|=v_0\exp{(\vphi/2)}$. Hence, the mass of the fermion: $m_\psi=g_\psi v_0\,e^{\vphi/2}$, is a point-dependent quantity as well. The resulting WIG theory of gravity and of the SMP preserves the Weyl symmetry even after $SU(2)\times U(1)$ symmetry breaking.\footnote{The Weyl invariance of the action for fermion fields in curved spacetime is demonstrated in Ref. \cite{cheng_prl_1988}.} This means that if CGR were the correct (classical) theory of gravity, Weyl invariance might be an actual symmetry of the laws of physics in our present universe.

At first sight the symmetry-breaking procedure driven by the Weyl invariant Lagrangian \eqref{higgs-lag} seems a bit confusing. Therefore, let us to briefly comment on symmetry breaking involving point dependent VEV. The self-interacting potential in the Lagrangian \eqref{higgs-lag}:

\bea V(|H|,\vphi)=\frac{\lambda'}{2}\left[|H|^2-v_0^2e^\vphi\right]^2,\label{higgs-pot}\eea is a function of the magnitude of the Higgs field $|H|$ and also of the gauge field $\vphi$. However, since $\vphi$ may be gauged away (it is not actually a dynamical degree of freedom), it may not be taken as an independent variable in what regards to the problem of finding the extrema of the potential, i. e., $V(|H|,\vphi)\rightarrow V(|H|)$. In other words, since in the neighborhood of each spacetime point $P_i$ the GR gauge with constant $\vphi_{0i}$ is singled out (see section \ref{sect-egr}), in general a different magnitude of the mass parameter $v_{0i}=v_0\exp(\vphi_{0i}/2)$ is set at each such spacetime point. Hence, the problem of finding maxima and minima of the potential \eqref{higgs-pot} at each $P_i$ is equivalent to the one for a single variable function, $V=V(|H|)$. The corresponding local VEV: $\left\langle H\right\rangle=\pm v_{0i}$, will be obviously a point-dependent quantity.

A perhaps more familiar example of such a point-dependent symmetry breaking is the symmetry restoration procedure at high temperatures \cite{linde-rpp}. Actually, at non-vanishing temperature $T$ all physically meaningful quantities are obtained not by vacuum averages in field theory but by Gibbs averages. In particular an equilibrium value of the field $\vphi(T)$ is given not by the minimum of the symmetry breaking potential $V(\vphi)$, but by the minimum of the free energy, $F(\vphi,T)\equiv V(\vphi,T)$, which at vanishing temperature $T=0$ coincides with $V(\vphi)$. In \cite{linde-rpp}, for instance, the following symmetry breaking potential: $$V(\vphi)=-\frac{\mu^2}{2}\vphi^2+\frac{\lambda}{4}\vphi^4,$$ is considered. At high temperatures $T\gg m$, where the mass of the scalar field is given by the following expression: $$m^2(\vphi)=\frac{dV}{d\vphi}=3\lambda\vphi^2-\mu^2,$$ the temperature-dependent potential is given by: $$V(\vphi,T)=V(\vphi)-\frac{\pi^2}{90}T^4+\frac{m^2(\vphi)}{24}T^2.$$ Hence, in a cosmological setting both: $\vphi=\vphi(t)$ and $T=T(a(t))$ ($a$ is the cosmological scale factor), are functions of the cosmic time and so are fields. Yet, there is not any motion equation for $T$, so that it is not considered as an independent field variable when the extrema of the above temperature-dependent potential are computed.


\section{Cosmology in the conformal general relativity}\label{sect-cosmo}

Let us to search for cosmological solutions of the motion equations in \eqref{winv-feq}. Here we consider a FRW background spacetime with flat spatial sections, with line-element: $ds^2=-dt^2+a^2(t)\delta_{ik}dx^idx^k$. For simplicity here we take the case with $\lambda=0$, so that the second term in the RHS of \eqref{wig-feq} vanishes. The case with $\lambda\neq 0$ will be considered separately. The Friedmann equation in \eqref{wig-feq} and the conservation equation \eqref{wig-conserv-eq} read,

\bea &&3\left(H+\frac{\dot\vphi}{2}\right)^2=\frac{\rho^{(w)}_m}{M^2_0},\nonumber\\
&&\dot\rho^{(w)}_m+3\gamma\left(H+\frac{\dot\vphi}{2}\right)\rho^{(w)}_m=0,\label{weyl-moteq}\eea or, after integrating the second equation above:

\bea 3\left(H+\frac{\dot\vphi}{2}\right)^2=\frac{M^4}{M^2_0}\frac{e^{-3\gamma\vphi/2}}{a^{3\gamma}},\label{weyl-cosmo-moteq}\eea where $M^4$ is an integration constant. In the above equations the matter contribution have been assumed in the form of a perfect barotropic fluid with energy density, $\rho^{(w)}_m\equiv e^{-\vphi}\rho_m$, and pressure, $p^{(w)}_m$, which obey the equation of state: $p^{(w)}_m=(\gamma-1)\rho^{(w)}_m$ ($\gamma$ is the barotropic index). 

Let us introduce the variable, $u\equiv a\exp\vphi/2$, so that \eqref{weyl-cosmo-moteq} reads: $$u^{3\gamma/2-1}\dot u=\sqrt\frac{M^4}{3M^2_0}.$$ Straightforward integration of the above equation yields:

\bea a(t)e^{\vphi(t)/2}=\alpha^\frac{2}{3\gamma}\left(t-t_0\right)^\frac{2}{3\gamma},\label{sol-1}\eea where $t_0$ is an (rescaled) integration constant and $$\alpha\equiv\frac{3\gamma}{2}\sqrt\frac{M^4}{3M^2_0}.$$ 

Given the gauge freedom inherent in the Weyl invariant theory of gravity \eqref{winv-action}, only the combination $u=ae^{\vphi/2}$ can be determined by the motion equations. We are free to choose either any $a(t)$ or any $\phi(t)$ we want. Assume, for instance, the following cosmic dynamics: $a(t)=a_0(t-t_0)^n$. In this case the dynamics of the scalar field is given by: $$\vphi(t)=\vphi_0+2\left(\frac{2}{3\gamma}-n\right)\ln(t-t_0),$$ where $\vphi_0\equiv 2\ln(\alpha^{2/3\gamma}/a_0)$. Notice that the choice, $n=2/3\gamma$, leads to the GR gauge where: $\vphi=\vphi_0$ and $a(t)=a_0(t-t_0)^{2/3\gamma}$. 

Another interesting situation is when the background fluid is vacuum: $\gamma=0$. In this case, integration of \eqref{weyl-cosmo-moteq} yields: $$a(t)e^{\vphi(t)/2}=C_0\exp\left(\sqrt\frac{M^4}{3M^2_0}\,t\right),$$ where $C_0$ is (the exponent of) an integration constant. Hence, the choice of the GR gauge where $\vphi=\vphi_0$, leads to de Sitter expansion: $$a(t)\propto e^{\sqrt{\Lambda/3}\,t},\;\;\Lambda=\frac{M^4}{M^2_0}.$$ 

Another particularly interesting gauge is when, $\vphi(t)=\vphi_0+2\sqrt{M^4/3M^2_0}\,t$. In this gauge, $a=C_0 e^{-\vphi_0/2}$, is a constant so that we get a static universe. This means that de Sitter expansion in GR where the space is taken to be Riemannian, is physically equivalent to a static universe where the affinity of the space is Weyl integrable. 

How it can be that the same experimental data that supports Riemannian-de Sitter expansion can serve as experimental evidence for a WIG static universe? The prompt answer is quite simple: While in the Riemannian de Sitter space the measured redshift is due to the expansion of the universe, in the WIG space the same redshift is due to variation of the units of measure during the cosmic evolution. Imagine that as a result of certain atomic transition a photon with frequency, $\hbar\omega\propto\Delta m_\text{Atom}\propto\exp(\vphi(t_p)/2)$, is emitted at some time $t_p$ in the past in a distant location, in a Weyl-integrable static space. The energy of the photon ($\hbar\omega$) when received at some latter time $t_0$, is not enough to excite the same atomic transition in a similar atom: $\Delta m^0_\text{Atom}\propto\exp(\vphi(t_0)/2)$. The measured redshift would be, $$\frac{\Delta m^0_\text{Atom}-\Delta m_\text{Atom}}{\hbar}\approx\sqrt\frac{M^4}{3M^2_0}\frac{(t_0-t_p)}{\hbar},$$ where we have assumed that $t_0, t_P\ll\sqrt{3M^2_0/M^4}$, i. e. we are considering time periods much more shorter that the lifetime of the universe $\sim H^{-1}_0\approx 10^{18}$ sec. 

Hence, even if the universe is static, in a Weyl-integrable space, due to the point-dependent property of the units of measure, there is room for the redshift (see the discussion in section \ref{sect-redshift}).


\subsection{Vacuum cosmology in the CGR}\label{subsect-vac-egr}

In standard GR the cosmological constant turns out to be a measure of the energy density of the vacuum. Anything that contributes to the energy density of vacuum acts like a cosmological constant. Then it makes sense to ignore the matter fields in \eqref{wig-feq} and to take into account the homogeneous $\lambda$-term, in order to look for the cosmological consequences of the energy density of vacuum. In this particular case ($\rho_m^{(w)}=0$, $\lambda\neq 0$) the field equation reads:

\bea H+\frac{\dot\vphi}{2}=\pm\sqrt\frac{\lambda}{3M^2_0}\,e^{\vphi/2}.\label{lambda-sol}\eea 

Below we shall consider separately interesting particular cases.


\subsubsection{de Sitter expansion}

For definiteness let us choose the positive branch and the particular case where $H=H_0$, i. e., de Sitter expansion. We get: $$\frac{\dot\vphi}{2}=\sqrt\frac{\lambda}{3M^2_0}\,e^{\vphi/2}-H_0,$$ which, after integration in quadratures yields; 

\bea \vphi(t)=\ln\left(\frac{3M^2_0}{\lambda}H_0\right)-2\ln\left(1-C_0\,e^{H_0 t}\right),\label{de-sitter-sol}\eea where $C_0$ is (the exponent of) an integration constant and $$-\infty<t\leq-\frac{\ln C_0}{H_0}\Rightarrow-\infty<\vphi<\infty.$$ 

It is seen that in the formal limit $t\rightarrow-\infty$, the Weyl gauge scalar, $$\vphi\rightarrow\ln\left(\frac{3M^2_0}{\lambda}\,H_0^2\right),$$ is basically a constant so that the GR gauge is approached in this limit. In other words, the WIG-based de Sitter expansion is naturally originated from GR de Sitter expansion. This is an example of a cosmic dynamics that continuously joints the GR gauge with another (non-GR) gauge.


\subsubsection{Static universe}

In this case, since $a(t)=a_0$ is a constant, then \eqref{lambda-sol} is written as: $$e^{-\vphi/2}\frac{\dot\vphi}{2}=\pm\sqrt{\frac{\lambda}{3M^2_0}},$$ or after integration in quadratures:

\bea \vphi_\pm(t)=-2\ln\sqrt{\frac{\lambda}{3M_0^2}}\left(t_0\mp t\right),\label{static-sol}\eea where $t_0$ is an integration constant.


\subsubsection{Solutions in terms of conformal time}

In general, if introduce the conformal time: $d\tau=dt/a(t)$, the Friedmann equation \eqref{lambda-sol} is written in the following form: 

\bea \frac{1}{u^2}\frac{du}{d\tau}=\pm\sqrt\frac{\lambda}{3M^2_0},\label{feq-c-time}\eea where, as before (see section \ref{sect-cosmo}), the variable $u\equiv ae^{\vphi/2},$ has been used. Straightforward integration of \eqref{feq-c-time} yields:

\bea a(\tau)\,e^{\vphi(\tau)/2}=\frac{\sqrt{3M^2_0/\lambda}}{\tau_0\mp\tau},\label{c-time-sol}\eea where $\tau_0$ is an integration constant. 

In the GR gauge ($\vphi=\vphi_0$) we get a singular cosmological evolution: $$a_\pm(\tau)=\frac{e^{-\vphi_0/2}\sqrt{3M^2_0/\lambda}}{\tau_0\mp\tau}.$$ Depending of the chosen branch of the solution we get different cosmic behavior.

\begin{itemize}

\item `$+$' branch. The universe undergoes cosmic expansion ($H=da/d\tau>0$) with a curvature singularity at $\tau=\tau_0$ ($-\infty<\tau\leq\tau_0$), where both $a$ and $H$ diverge.

\item `$-$' branch. The universe undergoes contraction ($H<0$) starting the evolution in a curvature singularity in the past: $\tau=-\tau_0$ ($-\tau_0\leq\tau<\infty$), where $H\rightarrow-\infty$ and $a\rightarrow\infty$, i. e., both blowup. 

\end{itemize}

Another outstanding gauge is the one where the universe is static ($a(\tau)=a_0$ is a constant). In this case the dynamics of the gauge scalar is given by: $$\vphi(\tau)=2\ln\sqrt\frac{3M^2_0}{a_0^2\lambda}-2\ln\left(\tau_0\mp\tau\right).$$

As seen the spectrum of possible cosmic behavior -- including the vacuum case -- is by far much more diverse in a WIG universe governed by the laws of CGR than in standard Riemannian GR. In the light of the gauge freedom inherent in the CGR, this diversity may seen problematic. Take for instance the vacuum case. The de Sitter solution ($H=H_0$) singles out a gauge where the Weyl scalar $\vphi$ evolves according to \eqref{de-sitter-sol}, meanwhile the static solution ($a=a_0$) singles out a different gauge where $\vphi$ evolves according to \eqref{static-sol}. Both gauges are physically equivalent. The question then is: which one of these gauges should we choose in order to appropriately describe the cosmic evolution of vacuum? Under the assumption that the question is well posed, the answer should not be trivial and we defer the discussion of this subject for further work.


\section{On the flatness, horizon and relict particle abundances puzzles}\label{sect-infl}

Cosmological inflation \cite{infl-guth, infl-albrecht, infl-linde-1, infl-linde-2, infl-linde-21, infl-linde-3, infl-linde-31, olive-phys-rept-1990, hyb-infl-linde, infl-reheat, liddle-prd-1994, barrow-prd-1995, lidsey-rmp-1997, assist-infl, infl-lyth, liddle-book, infl-n-gauss, infl-obs} is a mechanism that was developed in order to address several serious problems of the hot bigbang paradigm. Among these are the flatness, horizon and relict particle abundances problems. Let us to briefly explain how the mentioned issues arise in the framework of GR, which is one of the infinite gauges of CGR, and why these do not arise in other gauges of CGR-based cosmology.


\subsection{Flatness issue}

In order to understand what is the flatness problem about let us to write the GR equations of motion in a FRW background whose line element in co-moving spherical coordinates $t,r,\theta,\phi$, reads:

\bea ds^2_\text{GR}=-dt^2+\frac{a_\text{GR}^2(t)}{1-kr^2}dr^2+r^2a_\text{GR}^2(t)d\Omega^2,\label{frw-k}\eea where $a_\text{GR}(t)$ is the scale factor, $k=\pm 1,0$ is the curvature of the spatial sections and $d\Omega^2\equiv d\theta^2+\sin^2\theta d\phi^2$. For our purpose it is enough to write the GR Friedmann equation:

\bea 3H^2_\text{GR}+\frac{3k}{a^2_\text{GR}}=\frac{1}{M^2_0}\rho^\text{GR}_m,\label{fried-gr}\eea where $H_\text{GR}=\dot a_\text{GR}/a_\text{GR}$ is the GR Hubble rate, $M_0$ is the bare Planck mass and $\rho^\text{GR}_m$ is the energy density of the background matter fluid. In terms of the dimensionless (normalized) energy density: $\Omega^\text{GR}_m\equiv\rho^\text{GR}_m/3M^2_0H^2_\text{GR},$ the Friedmann equation \eqref{fried-gr} can be written in the following alternative way: 

\bea \chi_\text{GR}:=\Omega^\text{GR}_m-1=\frac{k}{a^2_\text{GR} H^2_\text{GR}},\label{ratio-gr}\eea where the quantity $\chi_\text{GR}$ measures the departure from spatial flatness. According to the standard analysis one founds in the bibliography -- see for instance the book \cite{liddle-book} -- it is used to assume that the contribution from the background matter is much more important than the curvature itself so that, to all purposes, the curvature contribution may be ignored. In this case, if take into account that $\rho^\text{GR}_m=M^4a^{-3\gamma}_\text{GR}$ ($\gamma$ is the barotyropic index of the fluid and $M$ is a constant with dimensions of mass), straightforward integration of the Friedmann equation yields $a_\text{GR}\propto t^{2/3\gamma}$ and $H_\text{GR}\propto t^{-1}$. Hence, $a_\text{GR}H_\text{GR}\propto t^{(2-3\gamma)/3\gamma}$, which leads to: $\chi_\text{GR}\propto k\,t^{(6\gamma-4)/3\gamma}$, with $\gamma>2/3$. This means that with the course of the cosmic expansion the departure of the background energy density from its critical value grows. For background radiation $\chi_\text{GR}\propto t$, while for dust: $\chi_\text{GR}\propto t^{2/3}$. This means that the classical initial conditions must be incredibly fine-tuned in such a way that the Universe were spatially flat to a high accuracy (for instance, at the time of nucleosynthesis $\chi_\text{GR}=\pm 10^{-16}$ \cite{lidsey-rmp-1997, liddle-book}). Besides, the fact that at present the Universe is still so close to spatially flat is also a problem because the Universe is so old \cite{olive-phys-rept-1990}: How the Universe lasted tens of billion years with $\Omega^\text{GR}_m$ still close to unity? Within the framework of GR-based cosmology the flatness issue may be explained if the Universe underwent an inflationary stage at very early times. In such a case $H_\text{GR}\approx H_0=$ const. $\Rightarrow a_\text{GR}=a_0\exp{(H_0 t)}$, $$\Omega^\text{GR}_m-1=\frac{k\,e^{-2H_0 t}}{a^2_0 H^2_0},$$ so that, with the course of the expansion any initial departure from spatial flatness is exponentially suppressed. 

The above analysis has been oversimplified since it was not based on the exact equations of motion given that the curvature term -- the one we are interested in here -- was ignored. The correct analysis should be based in the Friedmann equation \eqref{fried-gr}. In this case \eqref{ratio-gr} can be written in the following way \cite{olive-phys-rept-1990}:

\bea |\chi_\text{GR}|=\frac{1}{\left|\frac{M^4}{3kM^2_0}\,a_\text{GR}^{2-3\gamma}-1\right|}.\label{chi-gr}\eea From this expression -- which is the exact one -- it follows that as long as $\gamma>2/3$, with the course of the cosmic expansion, no matter how fast the expansion happens, $|\chi_\text{GR}|\rightarrow 1$, i. e., $\Omega^\text{GR}_m\rightarrow 0$, so that departure from spatial flatness $|\chi_\text{GR}|$ only increases with time from $|\chi_\text{GR}|\approx 0$ near of the bigbang, and up to $|\chi_\text{GR}|=1$ at late time. This is a consequence of the very unstable character of spatial flatness. It entails that working in the GR framework there is no other way to explain how any (even very small) initial departure from spatial flatness can be erased by the subsequent cosmic expansion, than setting $\gamma<2/3$. A particularly outstanding case is that of background vacuum energy density, for which $\gamma=0$. In this case we get that, at some time of the cosmic expansion, eventually: $$|\chi_\text{GR}|\approx\frac{3|k|M^2_0}{M^4}\,a_\text{GR}^{-2},$$ so that departure from spatial flatness decreases with the expansion. As a matter of fact, as the expansion proceeds the energy density of the curvature ($k/a^2_\text{GR}$) eventually dilutes and the constant energy density of vacuum $\rho_\text{vac}=M^4$ starts dominating to give rise to a period of de Sitter expansion: $H_\text{GR}=M^2/\sqrt{3}M_0$, so that any initial departure from spatial flatness is very quickly erased.

As seen, postulating the existence of matter with barotropic index $\gamma<2/3$ -- in particular vacuum or cosmological constant with $\gamma=0$ -- is the only way in which spatial flatness can be explained in the framework of general relativity. Recalling that GR is just one single (although peculiar) gauge within the CGR, one may wonder whether the additional degree of freedom inherent in the latter formalism could help to solve, or at least to alleviate the flatness issue.


\begin{figure*}[t!]
\includegraphics[width=5cm]{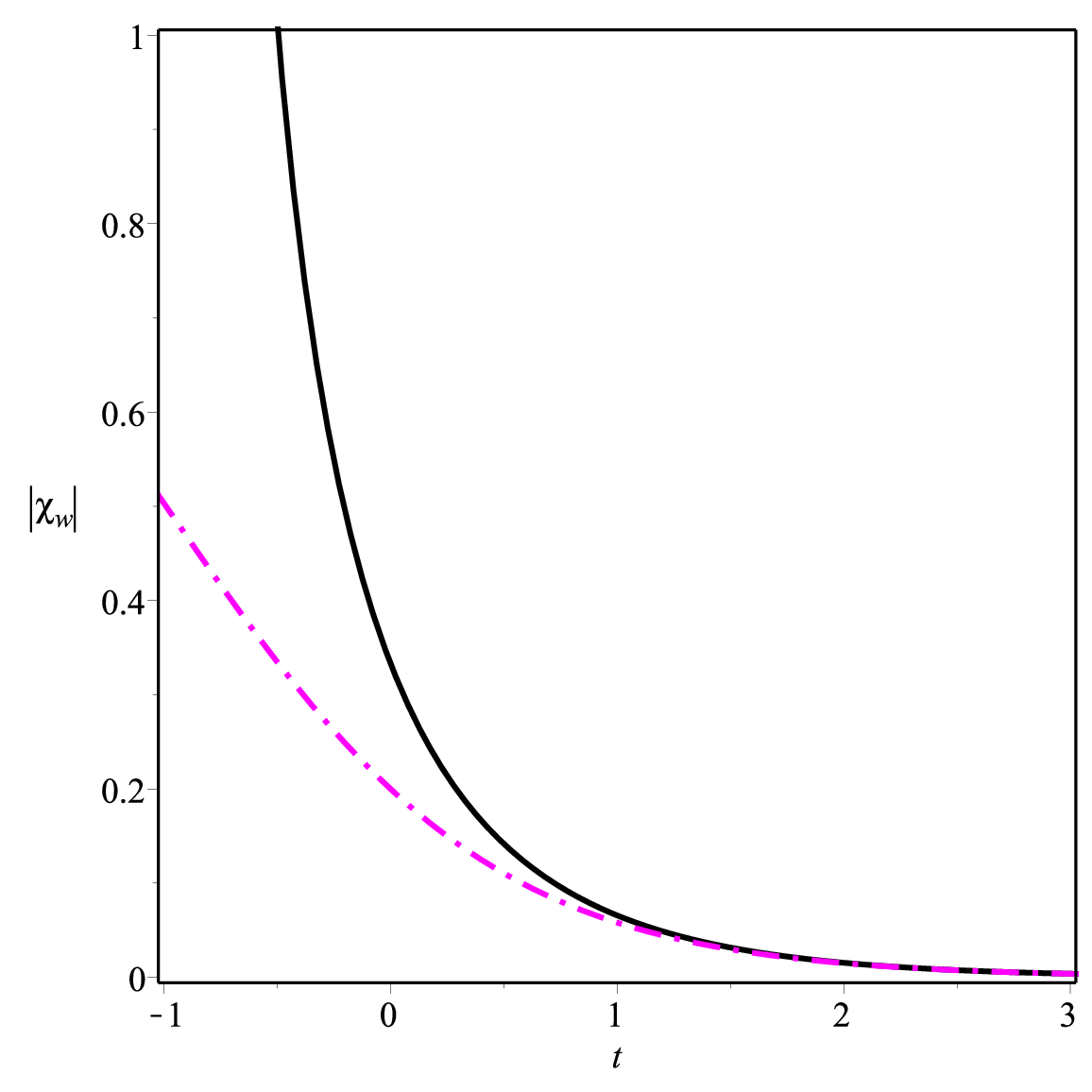}
\includegraphics[width=5cm]{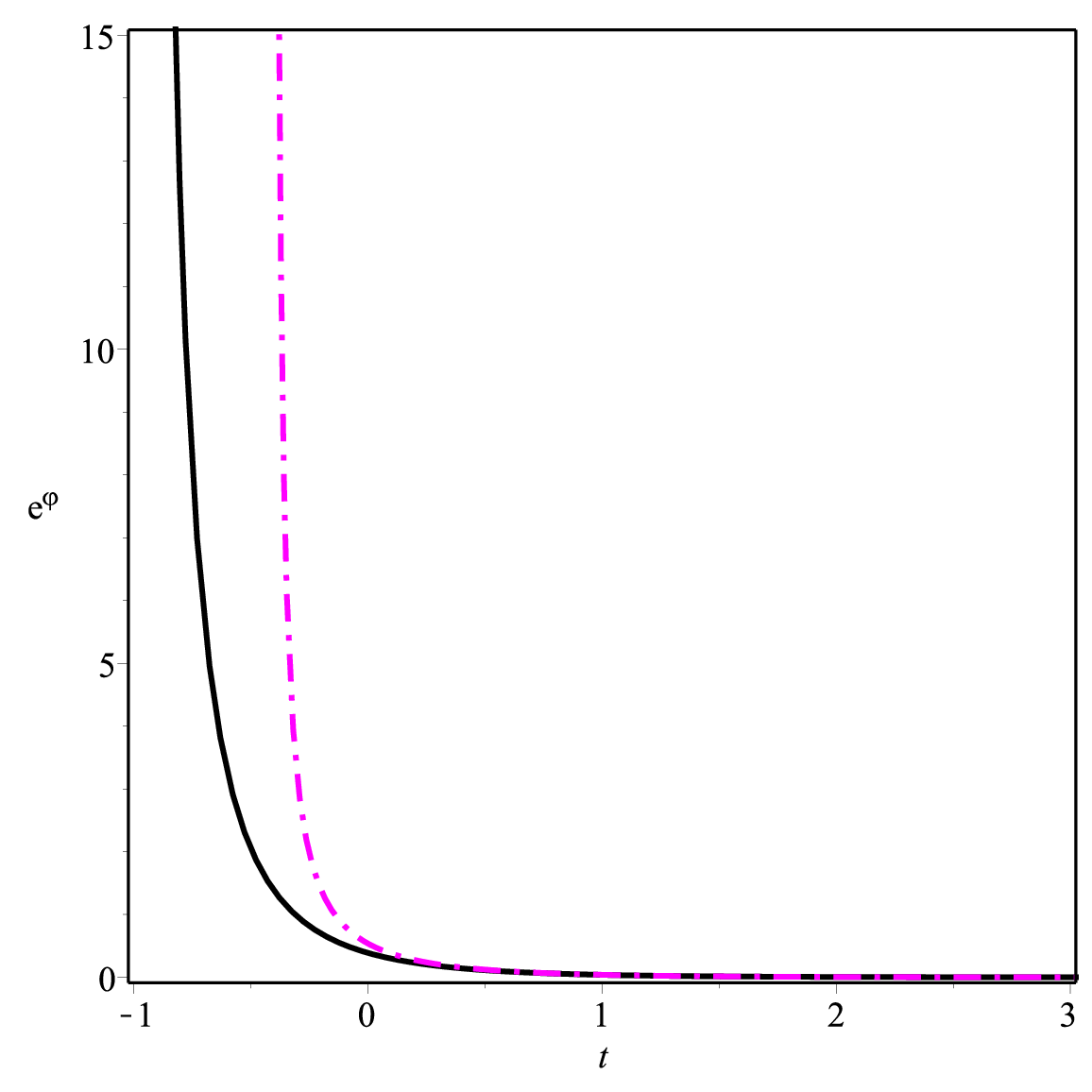}
\includegraphics[width=5cm]{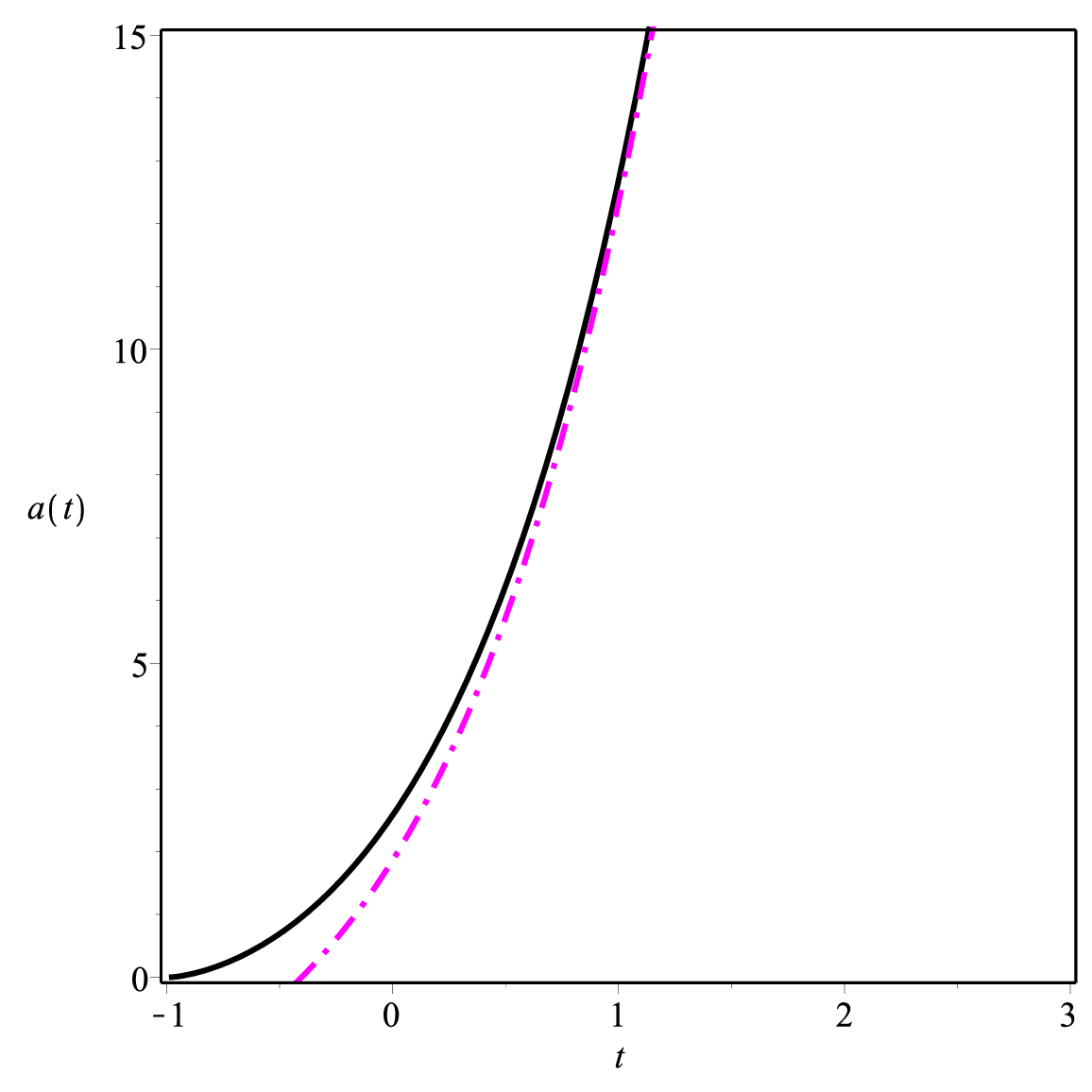}
\vspace{1.5cm}\caption{Plots of $|\chi_w|$ in \eqref{ratio-cgr-1} -- left -- of $\exp\vphi(t)$ in \eqref{ephi} -- middle -- and of the scale factor $a(t)$ in \eqref{scalef} -- right -- vs cosmic time $t$. Solid curves are for $k=1$ while dash-dots are for $k=-1$. The following values of the constant parameters: $\mu=0.7$ and $\alpha=2$, have been arbitrarily chosen.}\label{fig1}\end{figure*}



We shall show that indeed, due to gauge freedom in the framework of CGR-based cosmology, the flatness issue does not arise (but for the exception, of course, of the GR gauge). The Friedmann equation obtained from \eqref{wig-feq} (for simplicity we set $\lambda=0$) in a FRW spacetime with metric $$ds^2=-dt^2+\frac{a^2(t)}{1-kr^2}dr^2+r^2a^2(t)d\Omega^2,$$ reads:

\bea 3H^2_*+\frac{3k}{a^2}=\frac{1}{M^2_0}\,\rho^{(w)}_m,\label{wig-frw-k}\eea where $H_*=\dot a_*/a_*$ and $a_*:=a\exp(\vphi/2)$. Besides, the continuity equation is given by: $$\dot\rho^{(w)}_m+3\gamma H_*\rho^{(w)}_m=0\;\Rightarrow\;\rho^{(w)}_m=M^4 a_*^{-3\gamma}.$$ In terms of the dimensionless energy density $\Omega^{(w)}_m=\rho^{(w)}_m/3M^2_0H_*^2$, the Friedmann constraint \eqref{wig-frw-k} may be rewritten in the following way (compare with \eqref{ratio-gr}):

\bea \chi_w:=\Omega^{(w)}_m-1=\frac{k}{a^2H^2_*}=\frac{\left(a^\frac{3\gamma-2}{3\gamma}e^{\vphi/2}\right)^{3\gamma}}{\frac{M^4}{3kM^2_0}-\left(a^\frac{3\gamma-2}{3\gamma}e^{\vphi/2}\right)^{3\gamma}},\label{ratio-cgr}\eea where in the last equality in \eqref{ratio-cgr} we have substituted the motion equation \eqref{wig-frw-k}, so that \eqref{ratio-cgr} is an exact equation. From \eqref{ratio-cgr} it follows that, provided that $\exp(\vphi/2)$ decays faster than it grows $a^{1-2/3\gamma}$, there is not any issue with the spatial flatness. As a matter of fact, gauge freedom makes it possible to always find (at least) a gauge where the above condition is satisfied. As an illustration, let us to impose the following arbitrary constraint:\footnote{We can do this, i. e., to impose any condition we want on the combination $a(t)\exp\vphi(t)$ due, precisely, to gauge freedom.}

\bea a(t)e^{\vphi(t)}=e^{-\mu\,t},\label{arb-cond}\eea where $\mu\geq 0$ is a free parameter. Let us to imagine the Universe is filled with radiation, i. e., $\gamma=4/3$. The ratio \eqref{ratio-cgr} can then be written in the following way:

\bea |\chi_w|=\frac{3|k|M^2_0\,e^{-2\mu\,t}}{\left|M^4-3kM^2_0\,e^{-2\mu\,t}\right|}.\label{ratio-cgr-1}\eea Hence, under condition \eqref{arb-cond}, any initially existing departure from spatial flatness is very quickly erased by the cosmic expansion. If substitute the condition \eqref{arb-cond} into \eqref{wig-frw-k} one can readily integrate the Friedmann equation to obtain:

\bea e^{\vphi(t)}=\frac{\mu\,e^{-2\mu\,t}}{\alpha\ln\left|\frac{\sqrt{\alpha^2-k\,e^{-2\mu\,t}}+\alpha}{\sqrt{\alpha^2-k\,e^{-2\mu\,t}}-\alpha}\right|-2\sqrt{\alpha^2-k\,e^{-2\mu\,t}}},\label{ephi}\eea while for the scale factor,

\bea a(t)=\frac{1}{\mu}\,e^{\mu\,t}\left(\alpha\ln\left|\frac{\sqrt{\alpha^2-k\,e^{-2\mu\,t}}+\alpha}{\sqrt{\alpha^2-k\,e^{-2\mu\,t}}-\alpha}\right|-2\sqrt{\alpha^2-k\,e^{-2\mu\,t}}\right),\label{scalef}\eea where for compactness we have introduced the notation $\alpha\equiv M^2/\sqrt{3}M_0$. In FIG. \ref{fig1} the plots of $|\chi_w|$, $e^{\vphi(t)}$ and $a(t)$ vs the cosmic time $t$ are shown for chosen values of the free parameters.


\begin{figure*}[t!]
\includegraphics[width=5cm]{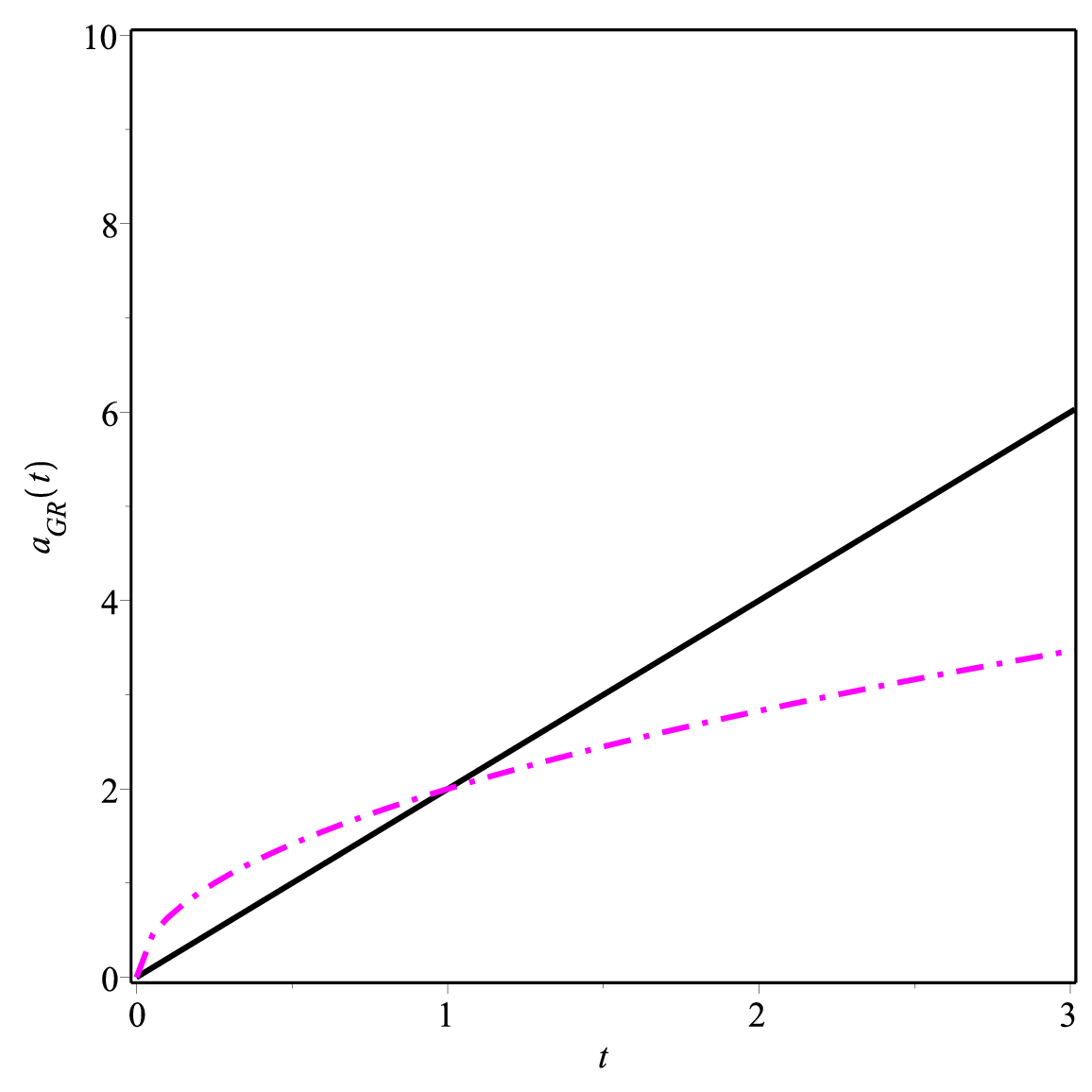}
\includegraphics[width=5cm]{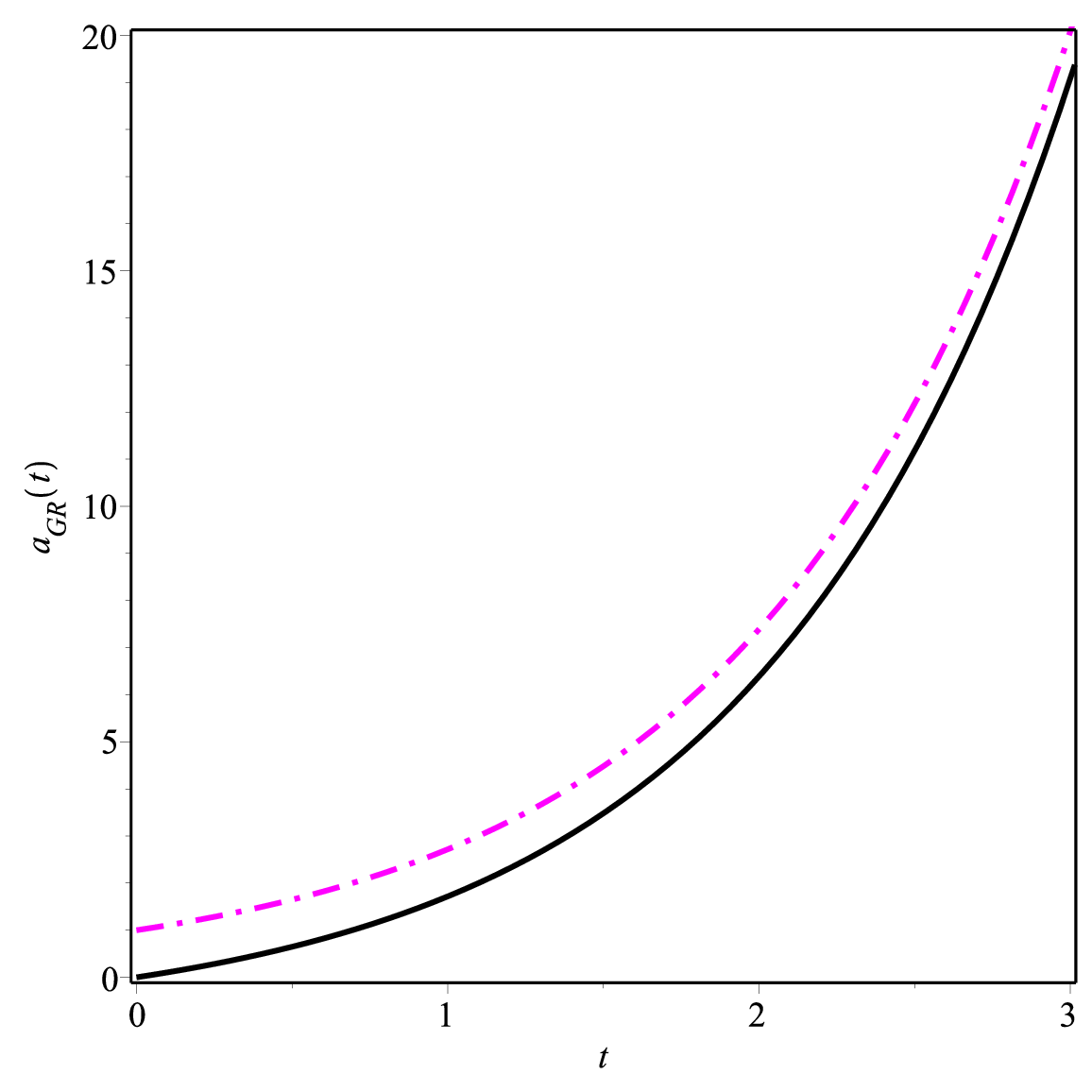}
\includegraphics[width=5cm]{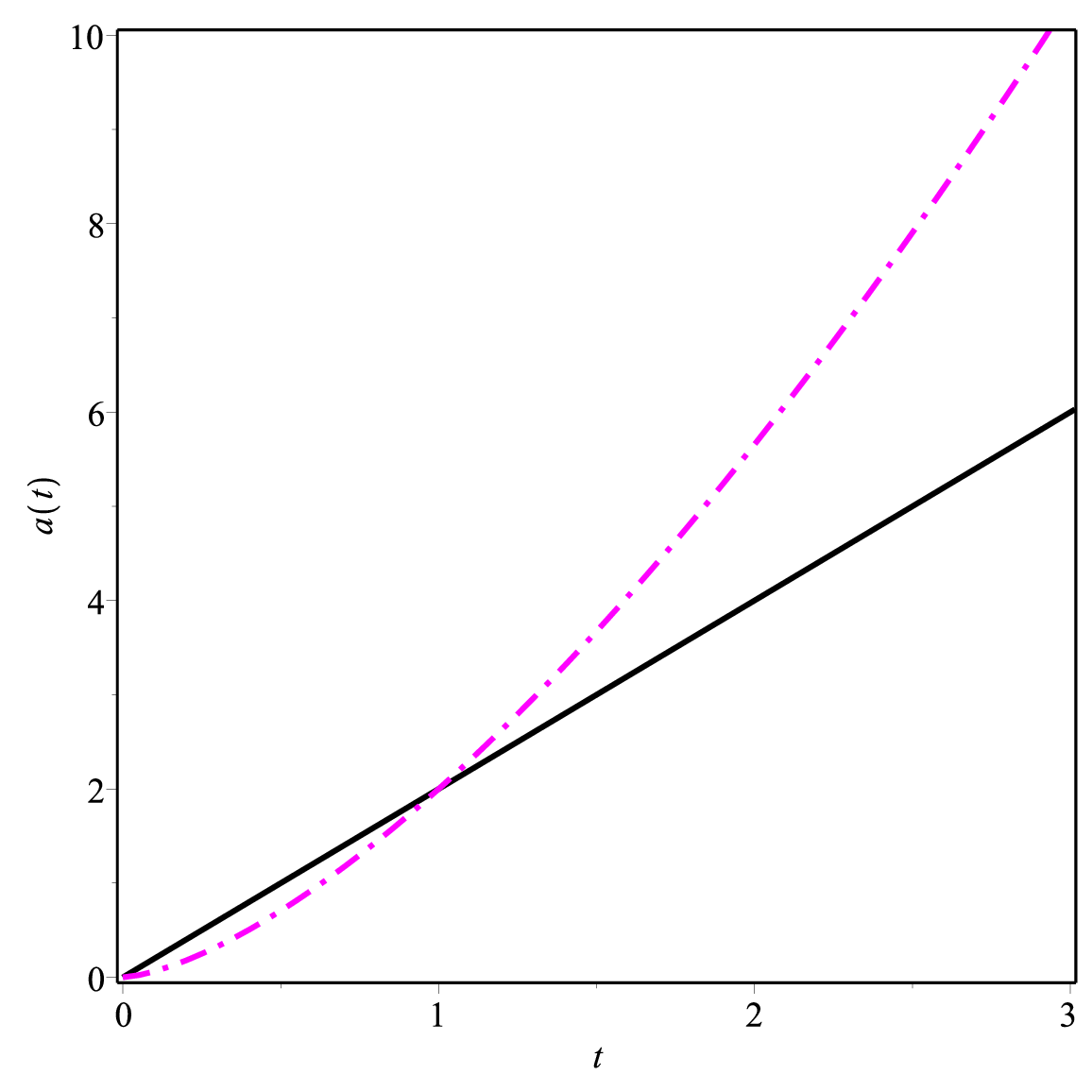}
\vspace{1.5cm}\caption{Plots of the scale factor -- dash-dots -- and of the distance to the causal horizon -- solid curves -- vs cosmic time $t$. The left-hand panel is for GR with background radiation where the horizon problem is evident, while the middle panel is for GR-de Sitter expansion where the horizon puzzle is settled. In the right-hand panel the plots are for an arbitrarily chosen gauge (different from the GR one) in the CGR where the horizon issue does not arise. }\label{fig2}\end{figure*}



\subsection{Horizon and relict particle abundances puzzles}

For the purpose of the present discussion it is enough to consider FRW spacetimes with flat spatial sections ($k=0$). The horizon problem arises because scales that originated outside of the causal horizon will eventually enter our past light cone and hence these will become part of our observable Universe \cite{olive-phys-rept-1990}. In consequence anisotropies are expected to be observed on large scales \cite{rindler, ellis}. 

Within the GR framework length scales grow as the scale factor: 

\bea a_\text{GR}=\left(\frac{3\gamma M^2}{2\sqrt{3}M_0}\right)^{2/3\gamma}\,t^{2/3\gamma},\label{agr}\eea where we have integrated \eqref{fried-gr} with $k=0$. Meanwhile, the causal horizon \cite{ellis} that is determined as the maximal physical distance light can travel from the co-moving position of an observer at some initial time to time $t$ \cite{olive-phys-rept-1990}: 

\bea d_H(t)=a_\text{GR}\int_0^t\frac{dt'}{a_\text{GR}(t')}=\frac{3\gamma\,t}{3\gamma-2}.\label{dh}\eea Because $\gamma>2/3$, the distance to the causal horizon grows faster than co-moving separations i. e., than the scale factor. The situation is illustrated in the left-hand panel of FIG. \ref{fig2} where the plots of $a_\text{GR}$ (dash-dots) and of $d_H$ (solid curve) vs the cosmic time $t$ are shown. It is seen that at very early times $t_\text{bb}$ (very close to the bigbang) and up to certain latter ``equality time'' $t_\text{eq}$ -- time at which the dash-dot and the solid curves meet again -- the curve representing $d_H(t)$ lies below of the curve for $a_\text{GR}(t)$. During this time interval, scales that at certain $t_\text{bb}\leq t\leq t_\text{eq}$ were located above the solid curve and below of the dash-dots, were not in causal contact with scales that were located below of the solid curve. After the equality time $t_\text{eq}$ those scales that were out of causal contact since the bigbang $t_\text{bb}$ and up to $t_\text{eq}$, enter the causal horizon so these can be seen by a co-moving observer. Inflation improves the situation since during this (nearly) de Sitter expansion period: $a_\text{GR}=H^{-1}_0\exp(H_0\,t)$, while $$d_H=\frac{1}{H_0}\left(e^{H_0\,t}-1\right).$$ The plots of $a_\text{GR}$ and of $d_H$ for this case are shown in the middle panel of FIG. \ref{fig2}. It is seen that scales that in the past were out of causal contact keep causally disconnected for all time.

As it was for the flatness problem, within the framework of CGR, due to gauge freedom, inflation is not required in order to explain the horizon issue. Actually, in the CGR massless particles are not affected by the WIG affinity of space, i. e., the photons paths are not affected by Weyl rescalings (see section \ref{sect-redshift}). This means that causality is not affected by the conformal transformation in the Weyl rescalings \eqref{weyl-t} so that the distance to the causal horizon \eqref{dh} still holds true in the CGR. On the other hand, since $a_*=a\exp(\vphi/2)$ is a guage invariant quantity, we have that:

\bea a\,e^{\vphi/2}=a_\text{GR}\;\Rightarrow\;a(t)=\left(\frac{3\gamma M^2}{2\sqrt{3}M_0}\right)^{2/3\gamma}\,t^{2/3\gamma}\,e^{-\vphi(t)/2},\label{acgr}\eea where we have substituted $a_\text{GR}$ from \eqref{agr} and, without loss of generality, we have set $\vphi_\text{GR}=\vphi_0=0$. Let us to specify a certain gauge by arbitrarily choosing: $\exp(-\vphi(t)/2)=t^n$ where $n$ is some non-negative power. In order to avoid the horizon issue it is enough that $n+3/2\gamma>1$, since in this case we will have that:

\bea a(t)=\left(\frac{3\gamma M^2}{2\sqrt{3}M_0}\right)^{2/3\gamma}\,t^{n+2/3\gamma},\;d_H(t)=\frac{3\gamma\,t}{3\gamma-2}.\label{adh-cgr}\eea This results in that co-moving separations grow faster than the distance to the causal horizon. The plots of $a(t)$ and $d_H(t)$ in \eqref{adh-cgr} vs the cosmic time $t$ are shown in the right-hand panel of FIG. \ref{fig2} for the case when $n=1$. It is seen that there are not scales formed outside of the causal horizon at the beginning of the expansion. Instead, with the course of the expansion there are scales that leave the causal horizon an can not be seen at any time in the future.

In what regards to the relict particles (magnetic monopoles, gravitinos, moduli fields, etc.) abundances, the question here is why the Universe is not dominated by these heavy relict particles at present? In order to explain why this is not so within GR-based cosmology it is again required the inflationary stage. Then, any initially existing amount of relict particles will be very quickly diluted $\rho^\text{GR}_\text{relict}\propto\exp(-3\gamma H_0\,t)$. Here, again, the analysis is not exact since de Sitter expansion is not a solution of the Einstein's equations with background matter and, besides, in general the state equation of the relict particles is very complex so that $P\neq (\gamma-1)\rho$. 

Within CGR-based cosmology the abundance of relict particles is not a problem as we shall see. Notice that $\rho_*=e^{-\vphi}\rho^{(w)}_\text{relict}$ is a gauge invariant quantity. Hence, $e^{-\vphi}\rho^{(w)}_\text{relict}=\rho^\text{GR}_\text{relict},$ where we have assumed that the equation of state $P=(\gamma-1)\rho$ holds true and that, in the GR gauge, $\vphi_\text{GR}=\vphi_0=0$. Since $\rho^\text{GR}_\text{relict}\propto a_\text{GR}^{-3\gamma}$ and given that the scale factor is given by \eqref{agr}, we have that $$\rho^{(w)}_\text{relict}=e^\vphi\rho^\text{GR}_\text{relict}\propto e^{\vphi(t)}t^{-2}=t^{-2(n+1)},$$ where in the last equality in the above chain we have assumed, as above, that $\exp(-\vphi(t)/2)=t^n$. From \eqref{agr} it follows that $t^{-2}\propto a_\text{GR}^{-3\gamma}$, so that the above relationship can be also rewritten in the following useful way:

\bea \rho^{(w)}_\text{relict}\propto a_\text{GR}^{-3\gamma(n+1)}.\label{density-cgr}\eea As seen the measured energy density of relict particles $\rho^{(w)}_\text{relict}$ in the given gauge of CGR decays much more quickly than the GR one: $\rho^{(w)}_\text{relict}\propto\left(\rho^\text{GR}_\text{relict}\right)^{n+1}$, where the choice of a specific value for the non-negative constant $n$ amounts to choosing a specific gauge within the CGR formalism ($n=0$ is for the GR gauge).


\section{On mass hierarchy and the cosmological constant problems}\label{sect-puzzles}

Among the puzzles of contemporary physics that still wait for a definitive resolution we may mention the mass hierarchy \cite{mass-h-1, mass-h-2, mass-h-3, mass-h-4, mass-h-5, mass-h-6} and the cosmological constant \cite{ccp-1, ccp-2, ccp-3, ccp-4, ccp-5, ccp-6, ccp-7} as the most outstanding problems. Fortunately, as we shall show below, these do not arise in the CGR. It is again gauge freedom the responsible for the avoidance of the mentioned problems. This is, definitively, one of the reasons why we believe that the CGR or some equivalent modification, may be taken seriously as an adequate alternative to plain GR.


\subsection{Mass hierarchy}\label{subsect-mass-h}

The mass hierarchy problem is about the required explanation to the large ratio between the electroweak scale $M_\text{EW}\sim 10^3$ GeV and the Planck scale $M_0\propto G_N^{-1/2}\sim 10^{18}$ GeV, 

\bea M_0\sim 10^{15}M_\text{EW}.\label{mass-h-eq}\eea 

There have been attempts at a resolution of this problem that are based in different approaches such as technicolor \cite{mass-h-1, mass-h-2, mass-h-3}, supersymmetry \cite{mass-h-4} or extra-dimensions \cite{mass-h-5, mass-h-6}. However, within standard GR no satisfactory resolution of this issue has been proposed. It is then very encouraging that, within the framework of WIG-based CGR, the mass hierarchy issue does not arise as we shall immediately show.

We first recall that in the CGR the measured value of the Newton's constant $G_N\propto M_0^{-2}$, is a true or bare constant in the sense that it is not affected by the Weyl rescalings (see section \ref{sect-newton-c}). Hence, the Planck scale, $M_0$, does not evolve in cosmic time, meanwhile the EW scale $M_\text{EW}$, being associated with the point-dependent EW symmetry breaking procedure described in section \ref{sect-weyl-smp}, in a cosmic setting is indeed a time-dependent quantity: $M_\text{EW}(t)\propto v_0\exp(\vphi(t)/2)$, where $v_0$ is the EW mass parameter. 

Let us suppose that at the time $t_*$ of the EW symmetry breaking phase, there was no hierarchy between the Planck and the EW mass scales: $$M_\text{EW}(t_*)\propto v_0\,e^{\vphi_*/2}\sim M_0,$$ where $\vphi_*=\vphi(t_*)$. At present time $t_0$, $M_\text{EW}(t_0)\propto v_0\,e^{\vphi_0/2}$, where $\vphi_0=\vphi(t_0)$. Hence, the ratio: 

\bea \frac{M_0}{M_\text{EW}(t_0)}\propto\frac{M_0}{v_0\,e^{\vphi_0/2}}=\frac{M_0\,e^{\Delta\vphi/2}}{M_\text{EW}(t_*)}\sim e^{\Delta\vphi/2},\label{ratio-today}\eea where $\Delta\vphi=\vphi_*-\vphi_0$. This means that the present observed ratio \eqref{mass-h-eq} is a natural consequence of a (not large) decrease in the magnitude of the gauge scalar: $\Delta\vphi\sim 70$, since the time of the EW symmetry breaking phase and up to the present age of the cosmic evolution.


\subsection{Vacuum energy and the cosmological constant}\label{subsect-vacuum}

If we trust quantum field theory up to scales of the order of the Planck mass, $M_0\sim 10^{18}$ GeV, then an approximate order of magnitude of the energy density of vacuum could be: 

\bea \rho_\text{vac}\propto M_0^4\sim 10^{72}\;\text{GeV}^4.\label{theor-estim}\eea The problem is that the observed value of the vacuum energy is bounded from above: $$|\rho^\text{obs}_\text{vac}|\leq 10^{-48}\;\text{GeV}^4,$$ so that a huge discrepancy between both quantities arises \cite{ccp-1, ccp-2, ccp-3, ccp-4, ccp-5, ccp-6, ccp-7}:

\bea \rho_\text{vac}\sim 10^{120}|\rho^\text{obs}_\text{vac}|.\label{ccp-eq}\eea 

In general Lorentz invariance leads to the following VEV of the vacuum energy-momentum tensor: $$\left\langle T^\text{vac}_{\mu\nu}\right\rangle=-\left\langle\rho_\text{vac}\right\rangle\,g_{\mu\nu}.$$ Hence, according to the WIG Einstein's field equation \eqref{wig-feq}, the quantity: $\rho^{(w)}_\lambda=\lambda\,\exp(\vphi)$, should be interpreted as the energy density of vacuum. 

Given that the gauge scalar $\vphi$ is a smooth (differentiable) function, it is intuitive that a small neighborhood of a given spacetime point can be always found, where $\vphi\approx\vphi_0$ is a constant. Hence, locally the GR gauge is singled out and the geometrical structure of the spacetime is well approached by Riemann geometry. This is required also by consistency with local Lorentz invariance. We face the following situation: On the one hand we can compute the magnitude of the vacuum energy density locally; $\rho_\text{vac}=\lambda\,\exp(\vphi_0),$ where $\vphi_0$ is the locally measured value of the gauge scalar, while, on the other hand, from the cosmological observations one may infer the magnitude of the vacuum energy density; $\rho^\text{obs}_\text{vac}=\lambda\,\exp(\vphi_\text{obs})$, where $\vphi_\text{obs}$ is the asymptotic cosmic value of the gauge scalar measured far from the region where the local measurements take place. Hence,

\bea \rho_\text{vac}=e^{\Delta\vphi}\rho^\text{obs}_\text{vac},\;\Delta\vphi=\vphi_0-\vphi_\text{obs},\label{ccp-ratio-eq}\eea so that, a not too large decrease of the magnitude of the cosmic gauge scalar as compared with its locally measured value: $\Delta\vphi\sim 276$, is enough to explain the large discrepancy between the theoretical estimate of the vacuum energy density \eqref{theor-estim} and its observed cosmic value \eqref{ccp-eq}.


\section{Discussion}\label{sect-discuss}

Weyl invariance of the physical laws -- in particular of the gravitational laws -- implies invariance of the action of the theory and, consequently, of the resulting equations of motion, under Weyl rescalings \eqref{weyl-t}: $$g_{\mu\nu}\rightarrow\Omega^{-2}g_{\mu\nu},\;\;\vphi\rightarrow\vphi+2\ln\Omega.$$ Additionally, it also requires invariance of the geometrical structure, in particular of the geodesics of the geometry, under these transformations. Otherwise, an issue with coupling of matter to gravity arises. The required invariance can be achieved if replace the Riemann geometrical structure of the background spacetime by the Weyl-integrable geometry where the units of measure are allowed to depend on the spacetime point. We end up with a theory where Weyl invariance is explicit in the action and, consequently, in the derived equations of motion \eqref{winv-feq}: 

\bea &&S=\int d^4x\sqrt{|g|}\left[\frac{M^2_0\,e^\vphi}{2} R^{(w)}-\lambda e^{2\vphi}+{\cal L}_m\right],\nonumber\\
&&G^{(w)}_{\mu\nu}=\frac{1}{M^2_0}\left[T^{(w,m)}_{\mu\nu}-\lambda e^\vphi g_{\mu\nu}\right],\label{unif-eq}\eea with the geometric quantities and operators defined in spacetimes with WIG affinity. 

Let us to show that, at least in appearance, the action in the above equation is equivalent to other Weyl invariant approaches found in the bibliography. Actually, given that the WIG-Ricci scalar can be written in terms of Riemannian quantities and operators \cite{novello, quiros_grg_2013}: $$R^{(w)}=R-\frac{3}{2}(\der\vphi)^2-3\Box\vphi,$$ it is straightforward to show that

\bea S=\frac{M^2_0}{2}\int d^4x\sqrt{|g|}\,e^\vphi R^{(w)}=\frac{M^2_0}{2}\int d^4x\sqrt{|g|}\,e^\vphi\left[R-\frac{3}{2}(\der\vphi)^2-3\Box\vphi\right]=\frac{M^2_0}{2}\int d^4x\sqrt{|g|}\,e^\vphi\left[R+\frac{3}{2}(\der\vphi)^2\right],\nonumber\eea where we have taken into account that: $$e^\vphi\Box\vphi=\nabla^\mu(e^\vphi\nabla_\mu\vphi)-e^\vphi(\der\vphi)^2,$$ and that the first term in the RHS above under the integral amounts to a boundary term that may be safely removed. Besides, if make the scalar field redefinition $e^\vphi=\phi^2$, we get: 

\bea S=\frac{M^2_0}{2}\int d^4x\sqrt{|g|}\,e^\vphi\left[R+\frac{3}{2}(\der\vphi)^2\right]=6M^2_0\int d^4x\sqrt{|g|}\left[\frac{\phi^2}{12}\,R+\frac{1}{2}(\der\vphi)^2\right].\nonumber\eea Hence, up to the factor $6M^2_0$, the action in \eqref{unif-eq} coincides with the action \eqref{deser-action}. The subtle difference is in the assumed geometrical properties of the background spaces.

Below we shall compare our setup with similar approaches in the bibliography in order to make clear what the differences are and how these differences impact the physical consequences of the theories.


\subsection{Weyl invariance in Riemann manifolds}

Let us to start with an approach which is popular when discussing on Weyl invariance of the kind we are discussing here \cite{bars, bars_1, bars_2, carrasco, quiros_1, jackiw, alpha, alpha_2, alpha_3, alpha_4}. The approach is based on the action \eqref{deser-action}: $$S=\int d^4x\sqrt{|g|}\left[\frac{\phi^2}{12}\,R+\frac{1}{2}(\der\phi)^2-\frac{\lambda}{12}\,\phi^4\right],$$ which, as we have just shown, coincides with the action in \eqref{unif-eq} up to the factor $6M^2_0$. However, in this case it is implicitly assumed that the background spacetimes are (pseudo)Riemannian, so that $\nabla_\lambda g_{\mu\nu}=0$, $\nabla_\lambda\sqrt{|g|}=0$, and so on. A clear difference of our Weyl invariant theory \eqref{unif-eq} when compared with the above theory is in the coupling of matter to gravity. In \eqref{unif-eq}, thanks to the WIG metricity condition: $$\nabla^{(w)}_\lambda g_{\mu\nu}=-\der_\lambda\vphi g_{\mu\nu}\Rightarrow\delta g_{\mu\nu}=-\delta\vphi g_{\mu\nu},$$ that leads to variations of the metric being affected by variations of the gauge scalar, any matter field can be coupled to gravity (see the derivation of the Einstein's and Klein-Gordon motion equations in appendix \ref{sect-app-b}). Meanwhile, in the theory \eqref{deser-action} only traceless matter fields may interact with gravity. Actually, the equations of motion that can be derived from \eqref{deser-action} with the addition of a minimally-coupled matter Lagrangian, read:

\bea &&G_{\mu\nu}=\frac{6}{\phi^2}\,T^{(m)}_{\mu\nu}-4\left[\frac{\der_\mu\phi}{\phi}\frac{\der_\nu\phi}{\phi}-\frac{1}{4}\,g_{\mu\nu}\left(\frac{\der\phi}{\phi}\right)^2\right]+\frac{2}{\phi}\left(\nabla_\mu\nabla_\nu-g_{\mu\nu}\Box\right)\phi,\nonumber\\
&&\Box\phi-\frac{\phi}{6}\,R=0.\label{deser-feq}\eea If one takes the trace of the first equation above one gets: $$\Box\phi-\frac{\phi}{6}\,R=\frac{1}{\phi}\,T_{(m)},$$ which coincides with the Klein-Gordon equation in \eqref{deser-feq} (which was derived through taking variation of the action \eqref{deser-action} with respect to the scalar field $\phi$) only if $T_{(m)}=0$. An alternative way to show the problem existing with the coupling of matter to gravity in the theory \eqref{deser-action} is by noting that, under a conformal transformation $g_{\mu\nu}\rightarrow\Omega^{-2}g_{\mu\nu}$, the continuity equation $$\nabla^\lambda T^{(m)}_{\lambda\mu}=0,$$ transforms into $$\nabla^\lambda T^{(m)}_{\lambda\mu}=-\der_\mu\left(\ln\Omega\right)\,T_{(m)}.$$ Hence, only for traceless matter the continuity equation is not transformed by the conformal transformation of the metric as required by conformal invariance. A similar analysis holds true for the theory \eqref{bars-action}: $$S=\int d^4x\sqrt{|g|}\left[\frac{\left(\phi^2-\sigma^2\right)}{12}\,R+\frac{1}{2}(\der\phi)^2-\frac{1}{2}(\der\sigma)^2\right].$$ 

In \cite{bars_1} the above action with $\sigma$ playing the role of the Higgs field, is complemented with a conformal invariant version of the Lagrangian of the SM (see equation (5) of that reference). The generation of masses and of the constants of nature in the resulting model is possible thanks to fixing the gauge field to an overall constant: $\phi=\phi_0$. The breakdown of Weyl symmetry through gauge fixing further triggers the Higgs mechanism and the generation of constant masses of particles with constant mass parameter $v\propto\phi_0$. This is in contrast to our model where the result of the `Higgs' mechanism is the generation of point-dependent masses of the SM particles through point-dependent mass parameter $v=v_0\exp(\vphi/2)$, so that the resulting physics remains Weyl invariant after the particles have acquired masses.


\subsection{Other similar approaches}

In reference \cite{scholz} a Weyl invariant theory of gravity defined over Weyl spaces is explored as an alternative to the mass generation scheme of the SMP. The action of this theory reads (we use our units and conventions):

\bea S=\int d^4x\sqrt{|g|}\left[-\frac{|D\phi|^2}{2}+\left(\pm\lambda|\phi|^4+\alpha R^{(w)}|\phi|^2\right)\right],\label{scholz-lag}\eea where $|D\phi|^2\equiv g^{\mu\nu}(D_\mu\phi)^\dag D_\nu\phi$, $D_\mu\phi=\der_\mu\phi-\vphi_\mu\phi$ is the gauge covariant derivative, $\alpha$ is a coupling constant and the geometric objects like the curvature scalar $R^{(w)}$ and the operators are defined in terms of a Weyl connection: 

\bea \Gamma^\alpha_{\mu\nu}=\{^\alpha_{\mu\nu}\}+\delta^\alpha_\mu\vphi_\nu+\delta^\alpha_\nu\vphi_\mu-g_{\mu\nu}\vphi^\alpha,\label{weyl-connection}\eea with $\vphi_\mu$ -- the Weyl gauge vector.\footnote{Do not confound the affine connection of a Weyl space \eqref{weyl-connection} with the WIG affine connection \eqref{weyl-aff-c} that corresponds to the particular case when the Weyl gauge vector $\vphi_\mu=\der_\mu\vphi$ equals to the gradient of a scalar.} The above action is invariant under the Weyl rescalings:

\bea g_{\mu\nu}\rightarrow\Omega^{-2}g_{\mu\nu},\;\phi\rightarrow\Omega\phi,\,\vphi_\mu\rightarrow\vphi_\mu+\frac{\der_\mu\Omega}{\Omega},\label{weyl-t'}\eea since, under \eqref{weyl-t'}, $|D\phi|^2\rightarrow\Omega^4|D\phi|^2$, and $R^{(w)}\rightarrow\Omega^2 R^{(w)}$ (it is assumed that the coupling constant $\alpha$ is unchanged by the Weyl rescalings). The differences of the theory \eqref{scholz-lag} with our setup are in the number and properties of the fields: In the latter theory, besides of the SM matter fields, there are the metric $g_{\mu\nu}$, the Weyl gauge vector $\vphi_\mu$ and the scalar field $\phi$. In our approach, in addition to the SM fields, there are only the metric and the gauge scalar.

An interesting version of WIG theory of gravity was explored in a series of papers \cite{romero_et_all, pucheu}. This theoretical setup is based on the action: 

\bea S=\int d^4x\sqrt{|g|}\left\{e^{-\vphi}\left[R^{(w)}-\omega(\vphi)(\der\vphi)^2\right]-V(\vphi)+{\cal L}_m\right\},\label{romero-action}\eea where $R^{(w)}$ is the WIG-Ricci scalar, $\omega=\omega(\vphi)$ is a coupling function, $\vphi$ is the Weyl scalar, i. e., the one in the definition of the WIG affine connection, and $V=V(\vphi)$ is the self-interaction potential. In this case $\vphi$ is a geometric scalar with direct physical implications. Since Weyl invariance is not required -- the above action is not Weyl invariant -- the Weyl scalar $\vphi$ can not be removed by a choice of gauge without physical consequences. 

The requirement of Weyl invariance of the laws of physics whose physical consequences have been explored in this paper have been advocated before in \cite{waldron, waldron-1}. However, we want to mention that in these references Weyl invariance is used to construct conformally invariant theories, as well as gauge invariant massless, massive and partially massless theories. These theories are constructed by means of the so called tractor calculus. The aim and scope of our paper has been by far more modest. Here we have explored the most salient physical consequences of the requirement of Weyl invariance in a simple version of conformal general relativity that was based on two postulates: i) the geometrical structure of the background space is that of Weyl-integrable geometry and ii) only the fundamental fields (and operators) are transformed by the Weyl resclaings. The bare constants such as the fundamental constants of nature, no matter whether these are dimensionless or dimensionful, are not transformed, i. e., these are Weyl invariant quantities.


\section{Conclusion}\label{sec-conclu}

The aim of this paper has been to challenge the standard point of view and to show that there can be an alternative for Weyl-invariance to survive to the coupling of any matter fields to gravity. This requires of a revision of the geometrical foundations of the GR theory including a modification of the coupling of the SMP fields to gravity. We have explored a Weyl invariant theory of gravity -- called as conformal general relativity -- which deals with spacetimes that are characterized by Weyl-integrable affinity. 

One of the most immediate consequences of assuming Weyl invariance of the laws of physics is the gauge freedom that adds to the coordinate freedom inherent in the known fundamental theories. It happens that instead of a definite solution of the motion equations -- like, for instance, in GR -- a whole set of physically equivalent conformal solutions arises. All of these solutions are consistent with the same experimental data and none if preferred over the others. Hence, in a Weyl invariant world -- with general relativity as a particular gauge -- the question about whether the experiment can determine the curvature of the space, can not be consistently settled: If the CGR were a correct (classical) theory of gravity, a space with curvature may be physically equivalent to a flat space with Weyl-integrable affinity, where the units of measure are point-dependent quantities.
 

Assuming that CGR is a correct description of the laws of gravity, amounts to assuming that any of the infinitely many physically equivalent gauges: $(g^{(i)}_{\mu\nu},\vphi_{(i)})$ (the index $i$ is a non-negative integer), $$g^{(i)}_{\mu\nu}=\Omega^2_{(i)} g^\text{GR}_{\mu\nu},\;\vphi_{(i)}=\vphi_\text{GR}-2\ln\Omega_{(i)},$$ where $g^\text{GR}_{\mu\nu}$ is any GR metric and $\vphi_\text{GR}=\vphi_0$ is a constant, can be also a correct description of the laws of gravity. Since under the above transformations the spacetime coincidences (the spacetime points) as well as the spacetime coordinates are unchanged, the experiment is not able to differentiate between these equivalent representations. Then, which one of the infinitely many equivalent descriptions to choose? Given that there is not any physical reason why we should prefer one over the others, the answer to this question is not trivial.

Yet, one may guest how to deal with the gauge freedom in a constructive way. Suppose that the gauge scalar is a smooth enough function, so that, locally the laws of gravity are well approached by GR. In order for the resulting picture to be consistent with the existing SMP and, at the same time, with the cosmological observational evidence, the flatness, horizon and relict particles abundances puzzles as well as the mass hierarchy and the cosmological constant issues, should find a natural explanation. As seen this is only achieved if assume that in the large cosmological scales GR is not anymore an acceptable approach and that the laws of CGR are the correct ones. Then one should look for a smooth function $\vphi(t,{\bf x})$ that can deal at the same time with the above mentioned puzzles in the manner explained in sections \ref{sect-infl} and \ref{sect-puzzles}. Anyway, we are left with an infinite set of such functions so that additional experimental evidence is required.

Another possible answer to the question about choosing a specific gauge is that, perhaps, we should change the way we state physically meaningful questions. In this regard the interesting question would not be, for instance, whether the expansion of the universe is accelerating or not, but, which one of the infinitely many physically equivalent conformal universes is the one we live in. Similarly, it is not of physical importance whether there is a black hole enclosing a curvature singularity or a singularity-free wormhole: one should choose the picture that allows going as far as possible with our calculations. Perhaps all of the infinitely many possibilities are potential descriptions and we are just ``trapped within one of them''.

In either case there is not observational evidence that favors GR over CGR (recall that GR is one specific gauge of CGR). The choice of one or another theory should rely on aesthetic arguments such as, for instance, whether the given theoretical setup is capable of explaining the given observational evidence in a way free of puzzles.



\section{acknowledgment}

The author thanks Erhard Scholz, Olivier Minazzoli, Serguei Odintsov and David Stefanyszyn for useful comments and the SNI of Mexico for support of this research.


\appendix



\section{Remarks on Weyl integrable geometry}\label{sect-app-a}

Weyl integrable geometry is a subclass of Weyl geometry which is free of the second clock effect \cite{smolin_npb_1979, zee_prd_1981, cheng_prl_1988, utiyama, novello, scholz_1, scholz_2, quiros_prd_2000, quiros_npb_2002, scholz, romero_ijmpa, romero_cqg, lobo, romero_et_all, pucheu, 2-clock-eff}. It represents the simplest deformation of Riemann geometry by allowing the length $l=\sqrt{g_{\mu\nu}l^\mu l^\nu}$ of vectors to change from point to point in space under parallel displacement: 

\bea dl=ldx^\mu\der_\mu\vphi\;\Rightarrow\;l=l_0\,e^{\vphi/2},\label{length}\eea where $\vphi$ is the Weyl gauge scalar or compensator field. This generalization of Riemann geometry rests on the following metricity condition:

\bea \nabla^{(w)}_\alpha g_{\mu\nu}=-\der_\alpha\vphi\,g_{\mu\nu}\;\Leftrightarrow\;\nabla^{(w)}_\alpha g^{\mu\nu}=\der_\alpha\vphi\,g^{\mu\nu},\label{met-cond}\eea where the supra(sub)-index ``$(w)$'' is for Weyl integrable quantities and operators. The above metricity condition leads to the affine connection of WIG to be given by:

\bea \Gamma^\sigma_{\mu\nu}\equiv\{^\sigma_{\mu\nu}\}+\frac{1}{2}\left(\delta^\sigma_\mu\der_\nu\vphi+\delta^\sigma_\nu\der_\mu\vphi-g_{\mu\nu}\der^\sigma\vphi\right).\label{weyl-aff-c}\eea It is very appealing that the above metricity condition \eqref{met-cond}, as well as the WIG affine connection \eqref{weyl-aff-c} are invariant under the Weyl rescalings \eqref{weyl-t}, $$g_{\mu\nu}\rightarrow\Omega^{-2}g_{\mu\nu},\;\;\vphi\rightarrow\vphi+2\ln\Omega.$$ The same applies to the WIG geodesics:

\bea \frac{d^2x^\mu}{ds^2}+\Gamma^\mu_{\sigma\lambda}\frac{dx^\sigma}{ds}\frac{dx^\lambda}{ds}-\frac{1}{2}\der_\lambda\vphi\frac{dx^\lambda}{ds}\frac{dx^\mu}{ds}=0.\label{weyl-geod-1}\eea 

A WIG space is given by the triad: $({\cal M},g_{\mu\nu},\vphi)$, where ${\cal M}$ represents the spacetime manifold. Hence, instead of a single WIG space one gets a whole class of them, that are related by the Weyl transformations \eqref{weyl-t}. This means that WIG spaces are the natural arena where to address Weyl invariance of the laws of gravity. Riemann geometry is retrieved in the particular gauge where $\vphi=\vphi_0$ is a constant. In this gauge the WIG affine connection \eqref{weyl-aff-c} transforms into the Christoffel symbols of the metric and the metricity condition \eqref{met-cond} is transformed into the condition: $\nabla_\alpha g_{\mu\nu}=0$, expressing that the metric is covariantly constant. In this case, under parallel transport, the length of vectors is unchanged.


\section{derivation of the motion equations in the CGR}\label{sect-app-b}

In order to derive the motion equations \eqref{winv-feq} from the action \eqref{winv-action}:

\bea S=\int d^4x\sqrt{|g|}\left[\frac{M^2_\text{Pl}(\vphi)}{2} R^{(w)}-\lambda\,e^{2\vphi}+{\cal L}_m\right],\nonumber\eea let us to split the integral into two pieces:

\bea &&S_1=\int d^4x\sqrt{|g|}\frac{M^2_\text{Pl}(\vphi)}{2} R^{(w)}=\frac{M^2_0}{2}\int d^4x\sqrt{|g|}\,e^\vphi R^{(w)},\nonumber\\
&&S_2=\int d^4x\sqrt{|g|}\left[-\lambda\,e^{2\vphi}+{\cal L}_m\right],\label{pieces}\eea respectively. Variation of the second piece with respect to the metric yields:

\bea \delta_g S_2=\frac{1}{2}\int d^4x\sqrt{|g|}\,\delta g^{\mu\nu}\left[\lambda\,e^{2\vphi}g_{\mu\nu}-T^{(m)}_{\mu\nu}\right],\label{ds1}\eea where we have taken into account the standard definition of the stress-energy tensor: $$T^{(m)}_{\mu\nu}=-\frac{2}{\sqrt{|g|}}\frac{\delta\left(\sqrt{|g|}{\cal L}_m\right)}{\delta g^{\mu\nu}}.$$ Meanwhile, if vary the piece $S_1$ above with respect to the metric we get:

\bea \delta_g S_1=\frac{M^2_0}{2}\int d^4x\sqrt{|g|}\,\delta g^{\mu\nu}e^\vphi\left[R^{(w)}_{\mu\nu}-\frac{1}{2}\,g_{\mu\nu}R^{(w)}\right]+\frac{M^2_0}{2}\int d^4x\sqrt{|g|}\,e^\vphi g^{\mu\nu}\delta R^{(w)}_{\mu\nu}.\label{var-main}\eea The second term in the RHS of the above equation is to be transformed. We have that $$\delta R^{(w)}_{\mu\nu}=\nabla^{(w)}_\lambda\left(\delta\Gamma^\lambda_{\mu\nu}\right)-\nabla^{(w)}_\nu\left(\delta\Gamma^\lambda_{\mu\lambda}\right),$$ hence,

\bea \frac{M^2_0}{2}\int d^4x\sqrt{|g|}\,e^\vphi g^{\mu\nu}\delta R^{(w)}_{\mu\nu}=\frac{M^2_0}{2}\int d^4x\sqrt{|g|}\,e^\vphi g^{\mu\nu}\left[\nabla^{(w)}_\lambda\left(\delta\Gamma^\lambda_{\mu\nu}\right)-\nabla^{(w)}_\nu\left(\delta\Gamma^\lambda_{\mu\lambda}\right)\right].\nonumber\eea Let us to show that the above term -- second term in the RHS of \eqref{var-main} -- vanishes. We have that,

\bea &&\int d^4x\sqrt{|g|}\,e^\vphi g^{\mu\nu}\nabla^{(w)}_\lambda\left(\delta\Gamma^\lambda_{\mu\nu}\right)=\int d^4x\nabla^{(w)}_\lambda\left(\sqrt{|g|}\,e^\vphi g^{\mu\nu}\delta\Gamma^\lambda_{\mu\nu}\right)\nonumber\\
&&\;\;\;\;\;\;\;\;\;\;\;\;\;\;\;\;\;\;\;\;\;\;\;\;\;\;\;\;\;\;\;\;\;\;\;\;\;\;\;\;\;\;\;\;\;\;\;\;-\int d^4x\sqrt{|g|}\,e^\vphi\delta\Gamma^\lambda_{\mu\nu}\left\{\left[\frac{\nabla^{(w)}_\lambda\sqrt{|g|}}{\sqrt{|g|}}+\der_\lambda\vphi\right]g^{\mu\nu}+\nabla^{(w)}_\lambda g^{\mu\nu}\right\},\nonumber\eea or, if remove the boundary term (first term in the RHS of the above equation) and take into account the metric compatibility condition \eqref{met-cond}, we obtain:

\bea \int d^4x\sqrt{|g|}\,e^\vphi g^{\mu\nu}\nabla^{(w)}_\lambda\left(\delta\Gamma^\lambda_{\mu\nu}\right)=-\int d^4x\sqrt{|g|}\,e^\vphi\delta\Gamma^\lambda_{\mu\nu}\left[-\der_\lambda\vphi\,g^{\mu\nu}+\nabla^{(w)}_\lambda g^{\mu\nu}\right]=0,\nonumber\eea where we have taken into consideration the following consequence of the metric compatibility condition \eqref{met-cond}: $$\nabla^{(w)}_\lambda\sqrt{|g|}=-2\der_\lambda\vphi\sqrt{|g|}\Rightarrow\frac{\nabla^{(w)}_\lambda\sqrt{|g|}}{\sqrt{|g|}}=-2\der_\lambda\vphi.$$ A similar derivation leads to: $$\int d^4x\sqrt{|g|}\,e^\vphi g^{\mu\nu}\nabla^{(w)}_\nu\left(\delta\Gamma^\lambda_{\mu\lambda}\right)=0.$$ This means that the term $$\propto\int d^4x\sqrt{|g|}\,e^\vphi g^{\mu\nu}\delta R^{(w)}_{\mu\nu},$$ in \eqref{var-main} may be safely ignored. 

After the above considerations we get that, varying the action \eqref{winv-action} with respect to the metric leads to: 

\bea \delta_g S=\frac{1}{2}\int d^4x\sqrt{|g|}\,\delta g^{\mu\nu}\left[M^2_\text{Pl}(\vphi) G^{(w)}_{\mu\nu}+\lambda\,e^{2\vphi}g_{\mu\nu}-T^{(m)}_{\mu\nu}\right],\label{equation}\eea from which the WIG-Einstein's motion equations -- first equation in \eqref{winv-feq} -- readily follow.

In order to obtain the motion equation for the gauge field $\vphi$ one has to vary the action \eqref{winv-action} with respect to $\vphi$. However, in WIG manifolds, thanks to the metric compatibility condition \eqref{met-cond}, variations of the metric and of the gauge field are not independent of each other. Actually, variation with respect to the gauge field $\vphi$ induces variation of the metric \eqref{delta-g}: $\delta g_{\mu\nu}=-g_{\mu\nu}\delta\vphi$ ($\delta g^{\mu\nu}=g^{\mu\nu}\delta\vphi$). Hence, by performing the straightforward substitution: $\delta\vphi\,g^{\mu\nu}\rightarrow\delta g^{\mu\nu}$, directly in \eqref{equation} we get that: 

\bea \delta_\vphi S=\frac{1}{2}\int d^4x\sqrt{|g|}\,\delta\vphi\left[-M^2_\text{Pl}(\vphi) R^{(w)}+4\lambda\,e^{2\vphi}-T_{(m)}\right],\label{equation-1}\eea from which the WIG Klein-Gordon equation in \eqref{winv-feq} follows. 

As clearly stated in the main text the Klein-Gordon equation for the gauge field is not an independent equation since it coincides with the trace of the WIG-Einstein's equations of motion. This is an inevitable consequence of the gauge freedom in connection with Weyl invariance.


\section{On the transformation properties of the fields and of the constants under Weyl rescalings}\label{sect-app-d}

An aspect of Weyl invariance that is worthy of discussion, is related with the transformation properties of the fields and of the constants of the theory under the Weyl rescalings \eqref{weyl-t}: $$g_{\mu\nu}\rightarrow\Omega^{-2}g_{\mu\nu},\;\;\vphi\rightarrow\vphi+2\ln\Omega.$$ In the bibliography one usually founds arguments that point to dimensionless constants as being unchanged by the Weyl rescalings, while dimensionful constants should be necessarily transformed under \eqref{weyl-t}. But then one also founds dimensionful constants, like the Planck constant $\hbar$, the speed of light $c$ and the electric charge of the electron $e$, that are assumed to be unchanged by the Weyl rescalings. The mentioned arguments are based on the interpretation of the conformal transformation of the metric as a transformation of the physical units. Alternatively, within the context of the gauge theories of the interactions, it is usually assumed that the constants of the theory, including the constant masses of fields, are unchanged by the gauge transformations. This is why, for instance, terms with the mass like in \eqref{symm-break}, are supposed to break the Weyl invariance.

Following the point of view on the conformal transformation as a units' transformation \cite{dicke-1962}, the statement that the dimensionful speed of light $c$ is unchanged by the Weyl rescalings \eqref{weyl-t} is readily understood if notice that (here, for simplicity we assume spacetimes with $g_{0i}=0$): $$c=\sqrt\frac{g_{ik}dx^i dx^k}{-g_{00}(dx^0)^2}.$$ Given that under \eqref{weyl-t} the coordinates are unchanged (see the related discussion in the introduction): $dx^\mu\rightarrow dx^\mu$, and that $g_{00}\rightarrow\Omega^{-2}g_{00}$ and $g_{ik}\rightarrow\Omega^{-2}g_{ik}$ transform in the same way, the Weyl invariance of the speed of light is immediate.\footnote{The argument is valid also in the more general situation where $g_{0i}\neq 0$ (this includes stationary metrics, etc.) and $$c_\pm=-\frac{g_{0i}dx^i}{g_{00}dx^0}\pm\sqrt{\left(\frac{g_{0i}dx^i}{g_{00}dx^0}\right)^2-\frac{g_{ik}dx^i dx^k}{g_{00}(dx^0)^2}}.$$} Unlike this, the argument that the Planck constant $\hbar$ is unchanged by the Weyl rescalings is more like an independent postulate. Actually, the usual explanation -- see, for instance, Ref. \cite{faraoni_prd_2007} -- is based on dimensional arguments and goes like this. The Planck constant has dimensions $[\hbar]=[ML^2T^{-1}]$, where $[M]$ stands for mass dimension, $[L]$ accounts for length dimension, while $[T]$ represents the time dimension. It is further assumed that the time and the length dimensions transform like the physical time and physical length, respectively. Hence, since the physical time interval $d\tau=\sqrt{-g_{00}}\,dx^0$ and the physical length element $dl=\sqrt{g_{ik}dx^i dx^k}$, transform in the same way under the Weyl rescalings: 

\bea d\tau\rightarrow\Omega^{-1}d\tau\Rightarrow [T]\rightarrow\Omega^{-1}[T],\;dl\rightarrow\Omega^{-1}dl\Rightarrow[L]\rightarrow\Omega^{-1}[L],\nonumber\eea the same is true for the time and the length dimensions. Provided that the mass dimension transforms like: $$m\rightarrow\Omega\,m\Rightarrow[M]\rightarrow\Omega[M],$$ any constant with the dimension of the Planck constant will be invariant under \eqref{weyl-t}. The fact that this last statement is an independent postulate can be seen if realize that the above transformation property of the Planck constant is a consequence of the assumption that the mass dimension is transformed like $[M]\rightarrow\Omega\,[M]$ under the Weyl rescalings. Under this assumption, for instance, given that $\phi\rightarrow\Omega\,\phi$ \eqref{scale-t}, a term like \eqref{symm-break} is invariant under \eqref{weyl-t}: $$\frac{1}{2}\int d^4x\sqrt{|g|}m^2\phi^2\rightarrow\frac{1}{2}\int d^4x\sqrt{|g|}m^2\phi^2.$$ Hence, under the assumption that the mass transforms like the inverse of the time/length dimension, the above mass term \eqref{symm-break} does not ensure Weyl symmetry breaking.

The above analysis was based on the assumption that the Weyl rescalings can be interpreted as transformations of the physical units in Dicke's sense \cite{dicke-1962}. However, in principle, one should be able to rescale the different physical units -- time and length, for instance -- in different ways, so that the Weyl transformations can be considered as very restricted transformations of units. The situation with the electric charge -- and with any other gauge charge -- is less clear and, assuming that it is not transformed by the Weyl rescalings is, definitively, an independent assumption. 

In the understanding that the Weyl rescalings can be identified with the transformations of units, it happens that the physical units, being constants in one formulation of the theory, are transformed into point-dependent units under the Weyl rescalings. This is what is called as running units in \cite{faraoni_prd_2007}. But, as we shall see, point-dependent units are not compatible with the affine properties of Riemannian spaces. In order to demonstrate this statement we shall consider two identical physical systems $A$ and $B$, that are located at different spacetime points. Let us focus in the measurement of an (coordinate) invariant quantity such as, for instance, the rest mass of the system: $M=\sqrt{g_{\mu\nu}P^\mu P^\nu}$, where $P^\mu$ is its 4-momentum. Hence, $M_A$ is the rest mass of the system $A$, while $M_B$ is the mass of the identical system $B$ that is located at a different spacetime point. Take another physical system to be the standard of measurement, where $m=\sqrt{g_{\mu\nu}p^\mu p^\nu}$ is the mass of the standard of measurement ($p^\mu$ is its 4-momentum). In other words, $m$ represents the standard mass unit. In order to measure the mass of $A$ the standard of measurement is to be parallelly transported to the point where $A$ is located and, then, the quantities $M_A$ and $m$ are to be compared: $M_A=\mu_A m$, where the dimensionless number $\mu_A$ is the result of the measurement. Then, in order to measure the mass of $B$, the standard of measurement should be parallelly transported to the point where $B$ is located and the above measurement procedure is repeated. We get that $M_B=\mu_B m$. It is obvious that, since $A$ and $B$ are assumed to be identical, then $\mu_A=\mu_B$. This means that the only way in which the quantity $M$ can vary from point to point in spacetime, i. e., that $M_A\neq M_B$, is that the standard unit of mass $m$ itself be a point-dependent quantity. 

However, it is a very well known fact that in Riemannian space the length of vectors does not change under parallel transport, i. e., $\nabla_\mu m=0\Rightarrow m=m_0$, is a constant over the spacetime (compare with \eqref{nabla-mass-wig} below). This is reminiscent of the metricity condition or metric compatibility of Riemann spaces, according to which the spacetime metric is covariantly constant: $\nabla_\sigma g_{\mu\nu}=0$ (compare with the metricity condition of WIG spaces \eqref{met-cond} in the appendix \ref{sect-app-a}). Hence, in Riemann spaces the (coordinate) invariant quantities, such like the rest mass, can not be point-dependent quantities. Yet one might assume that the rest mass is a point-dependent quantity, $m=m(x)$, but this should be an independent postulate of the theory that would have geometrical consequences. For instance, timelike particles with point-dependent mass $m(x)$ do not follow geodesics of the Riemann manifold due to the effect of an universal fifth-force. If these geometrical consequences are not taken into account, this can lead to inconsistent inferences made on the basis of the measurement process.

The situation is drastically changed if assume WIG to be the geometrical structure of the underlying spacetime. In a WIG space -- see the appendix \ref{sect-app-a} -- the length of vectors is allowed to vary from point to point under parallel transport: 

\bea \nabla_\mu m=\frac{1}{2}m\nabla_\mu\vphi\Rightarrow m=m_0\exp(\vphi/2),\label{nabla-mass-wig}\eea where $\vphi=\vphi(x)$ is the WIG gauge scalar. In this case the result of the measurement procedure explained above would yield to 

\bea m(\vphi)=\mu m_0\,e^{\vphi/2}=M_0\,e^{\vphi/2},\label{measured-mass}\eea i. e., $m(\vphi)$ would be a point-dependent quantity so that, for the above two identical physical systems $A$ and $B$: $m(\vphi_A)\neq m(\vphi_B)$, unless $\vphi=\vphi_0$ is a constant. Notice that, since under the Weyl rescalings \eqref{weyl-t}: $\vphi\rightarrow\vphi+2\ln\Omega$, then the point-dependent mass $m(\vphi)$ in \eqref{measured-mass} transforms under the Weyl rescalings \eqref{weyl-t} like: $m(\vphi)\rightarrow\Omega\,m(\vphi)$, where it is implicit that the (dimensionful) constant mass parameter $M_0$ is unchanged by \eqref{weyl-t}.

What if, following the spirit of the gauge theories of the fundamental interactions, look at the Weyl rescalings not as transformations of units but just as gauge transformations of the fields? In this case, assuming that under the Weyl rescalings the dimensionless field $\vphi$ transforms like: $\vphi\rightarrow\vphi+2\ln\Omega$, as in \eqref{weyl-t}, would entail that the also dimensionless quantity, $\exp\vphi/2\rightarrow\Omega\exp\vphi/2$, would be transformed in a different way. Hence, the combination $\exp\vphi\,R\rightarrow\Omega^4\exp\vphi\,R$, so that the action, $\int d^4x\sqrt{|g|}\exp\vphi\,R$, is not transformed by the Weyl rescalings. 

It follows from the above discussion that there are dimensionless quantities, like $\exp\vphi$, that are transformed by the Weyl rescalings, while dimensionful constants, such as $M_0$ in \eqref{measured-mass}, may be invariant under \eqref{weyl-t}. This means that the transformation properties of the basic quantities under the Weyl rescalings are to be postulated while the derived quantities are transformed accordingly. Under this reading the statement frequently found in the bibliography on scale invariance: ``Weyl symmetry does not allow any dimensionful parameters in the action...'' were correct only if implicitly assume Riemann geometry to govern the affine properties of spacetime, i. e., if these parameters were not supposed to vary from point to point in spacetime, but not in general.


\end{document}